\documentclass{egpubl}
\usepackage{eg2024}

\usepackage[T1]{fontenc}
\usepackage{dfadobe}  

\usepackage{cite}  % comment out for biblatex with backend=biber
\BibtexOrBiblatex
\electronicVersion
\PrintedOrElectronic
\ifpdf \usepackage[pdftex]{graphicx} \pdfcompresslevel=9
\else \usepackage[dvips]{graphicx} \fi

\usepackage{egweblnk} 
\title[Computational Smocking through Fabric-Thread Interaction]%
      {Computational Smocking through Fabric-Thread Interaction}

\author[N. Zhou et al.]
{\parbox{\textwidth}{\centering           
        Ningfeng Zhou\orcid{0009-0000-0108-6078}\ \ \ \ \ 
        Jing Ren\orcid{0000-0003-3114-3517} \ \ \ \ \ 
        Olga Sorkine-Hornung\orcid{0000-0002-8089-3974} 
         }
         \\
{\parbox{\textwidth}{\centering ETH Zurich, Switzerland}}}

\usepackage{booktabs} % For formal tables
\usepackage{units} % for nice formatting of values with units

\usepackage{algorithm}
\usepackage{algpseudocode}

\usepackage{hhline}
\usepackage{overpic}
\usepackage{bbold} % all ten digits with \mathbb
\usepackage{amsfonts} 
\usepackage{amsmath} 
\usepackage{pgfplots}
\usepackage{pgf,tikz}
\usepackage{tkz-euclide}
\usepackage{bm}
\usepackage{wrapfig}
\usepackage{multirow}
\usepackage{mathtools}
\usepackage{enumerate}
\usepackage{enumitem}
\usepackage{verbatim}

\usepackage{dsfont}
\newcommand{\R}{\mathds{R}} 
\newcommand{\D}{\mathds{D}} 
\newcommand{\proj}{\mathds{P}}

\newcommand{\V}{\mathcal{V}}
\newcommand{\E}{\mathcal{E}}

\renewcommand{\L}{\mathcal{L}}
\renewcommand{\P}{\mathcal{P}}

\DeclareMathOperator*{\argmin}{arg\,min}

\newcommand{\matr}[1]{\mathit{\mathbf{#1}}}% matrix

\usepackage{soul}

\usepackage{listings}
\usepackage{color, colortbl} %red, green, blue, yellow, cyan, magenta, black, white
\definecolor{mygreen}{RGB}{28,172,0} % color values Red, Green, Blue
\definecolor{mylilas}{RGB}{170,55,241}
\definecolor{ffzzcc}{rgb}{1,0.6,0.8}
\definecolor{mytbcol}{RGB}{175,227,246}
\definecolor{mypink}{RGB}{254, 197, 187}
\definecolor{tabyellow}{HTML}{7AAAF2} %yellow: ffd60a

\usepackage{lipsum}

\DeclareFontFamily{U}{mathx}{\hyphenchar\font45}
\DeclareFontShape{U}{mathx}{m}{n}{<-> mathx10}{}
\DeclareSymbolFont{mathx}{U}{mathx}{m}{n}
\usepackage{tipa}
\UndeclareTextCommand{\!}{T3}
\DeclareTextCommand{\tipaEXCLAM}{T3}{}
\DeclareRobustCommand{\!}{%
  \ifmmode\mskip-\thinmuskip\else\expandafter\tipaEXCLAM\fi
}

\usepackage{booktabs}       % toprules
\usepackage{mathrsfs}
\usetikzlibrary{arrows}
\usetikzlibrary{matrix}
\usetikzlibrary{positioning,calc,fadings}
\usetikzlibrary{backgrounds, fit}
\pgfplotsset{compat=1.14}
\usepackage{pgfplotstable}

\definecolor{mypurple}{HTML}{AB8AE0}
\definecolor{mygreen}{HTML}{38BBA1}
\definecolor{myorange}{HTML}{F69572}
\definecolor{myblue}{HTML}{6378B1}
\definecolor{myred}{HTML}{B13647}
\newcommand{\arap}{{\textsc{arap}}}

\newcommand{\cipc}{{\textsc{C-IPC}}}

\newif\ifdraft

\drafttrue

\ifdraft
\newcommand{\JR}[1]{{\color{red}[\textbf{JR:} #1]}}
\newcommand{\NF}[1]{{\color{orange}[\textbf{NF:} #1]}}
\newcommand{\OSH}[1]{{\color{magenta}[\textbf{OSH:} #1]}}

\newcommand{\new}[1]{{\color{black}#1}}

\newcommand{\todo}[1]{{\color{red}[\textbf{TODO:} #1]}}

\else
\newcommand{\JR}[1]{}
\newcommand{\NF}[1]{}
\newcommand{\OSH}[1]{}

\newcommand{\todo}[1]{}

\fi

\newcommand{\secref}[1]{Sec.~\ref{#1}}

\newcommand{\figref}[1]{Fig.~\ref{#1}}

\newcommand{\figreflist}[2]{Figures~\ref{#1} and~\ref{#2}}

\newcommand{\eqnref}[1]{Eq.~(\ref{#1})}

\newcommand{\Vs}{\textcolor{myorange}{\mathcal{V}_s}}
\newcommand{\Vf}{\textcolor{mygreen}{\mathcal{V}_f}}
\newcommand{\Vb}{\textcolor{mypurple}{\mathcal{V}_b}}
\newcommand{\Vp}{\textcolor{gray}{\mathcal{V}_p}}
\newcommand{\Es}{\E_s}
\newcommand{\Ef}{\E_f}

\usepackage{pifont}% http://ctan.org/pkg/pifont
\newcommand{\cmark}{\ding{52}}%
\newcommand{\xmark}{{\ding{56}}}%
\usepackage{accsupp}
\newcommand{\checked}{%
  \BeginAccSupp{method=hex,unicode}%
  \text{\rlap{\hskip 2pt \raisebox{0.5pt}{\normalsize{$\backprime$}}}\cmark}%
  \EndAccSupp{}%
}
\begin{document}

% uncomment for using teaser
% \teaser{
%  \includegraphics[width=\linewidth]{figures/teaser.pdf}
%  \centering
%   \caption{New EG Logo}
% \label{fig:teaser}
% }

\maketitle
%-------------------------------------------------------------------------
\begin{abstract}
We formalize Italian smocking, an intricate embroidery technique that gathers flat fabric into pleats along meandering lines of stitches, resulting in pleats that fold and gather where the stitching veers.
In contrast to English smocking, characterized by colorful stitches decorating uniformly shaped pleats, and Canadian smocking, which uses localized knots to form voluminous pleats, Italian smocking permits the fabric to move freely along the stitched threads following curved paths, resulting in complex and unpredictable pleats with highly diverse, irregular structures, achieved simply by pulling on the threads.
We introduce a novel method for digital previewing of Italian smocking results, given the thread stitching path as input. Our method uses a coarse-grained mass-spring system to simulate the interaction between \new{the }threads and the fabric. This configuration guides the fine-level fabric deformation through an adaptation of the \new{state-of-the-art simulator, C-IPC \cite{li2021CIPC}}.
Our method models the general problem of fabric-thread interaction and can be readily adapted to preview Canadian smocking as well.
We compare our results to baseline approaches and physical fabrications to demonstrate the accuracy of our method.
%

%-------------------------------------------------------------------------
%  ACM CCS 1998
%  (see https://www.acm.org/publications/computing-classification-system/1998)
% \begin{classification} % according to https://www.acm.org/publications/computing-classification-system/1998
% \CCScat{Computer Graphics}{I.3.3}{Picture/Image Generation}{Line and curve generation}
% \end{classification}
%-------------------------------------------------------------------------
%  ACM CCS 2012
%   (see https://www.acm.org/publications/class-2012)
%The tool at \url{http://dl.acm.org/ccs.cfm} can be used to generate
% CCS codes.
%Example:


\end{abstract}  
%-------------------------------------------------------------------------
\section{Introduction}

Embroidery, one of the oldest forms of art, produces exquisite decorations through the interplay of fabric and threads, and continues to captivate researchers in visual computing~\cite{efrat2016hybrid,ma2022multilayered,ren2023smocking,zhenyuan2023directionality}.
A recent paper~\cite{ren2023smocking} delves into Canadian smocking, an embroidery technique renowned for its intricate and voluminous pleats, whose geometric shape is difficult to predict by looking at the smocking pattern alone.
During the fabrication of Canadian smocking, multiple points along the same stitching line are gathered and secured with a knot. To compute a digital preview of a smocked result, Ren et al.~\cite{ren2023smocking} cast Canadian smocking as a graph embedding problem, merging multiple graph nodes into a single one, as fabric thickness has a negligible effect in this technique.
However, this method is not applicable to \emph{Italian smocking}, a different traditional smocking technique, in which fabric thickness significantly influences pleat formation and cannot be disregarded.

\begin{figure}[!t]
    \centering
    \begin{overpic}[trim=0cm 0cm 0cm -1cm,clip,width=1\linewidth,grid=false]{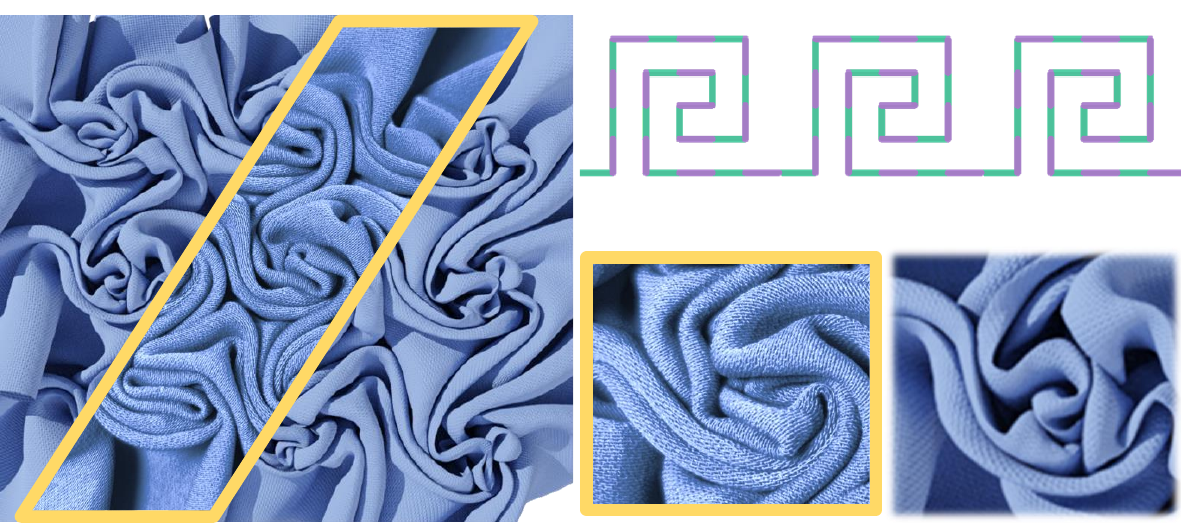}
    \put(57,45){\footnotesize \itshape smocking pattern (partial)}
    \put(8,45){\footnotesize \itshape our result \& fabrication}
    \put(51,25){\footnotesize \itshape zoom-in (fabric)}
    \put(78,25){\footnotesize \itshape zoom-in (ours)}
    \end{overpic}\vspace{-6pt}
    \caption{Our simulated result and physical fabrication (highlighted in yellow) for an Italian smocking pattern.}\label{fig:teaser}\vspace{-15pt}
\end{figure}
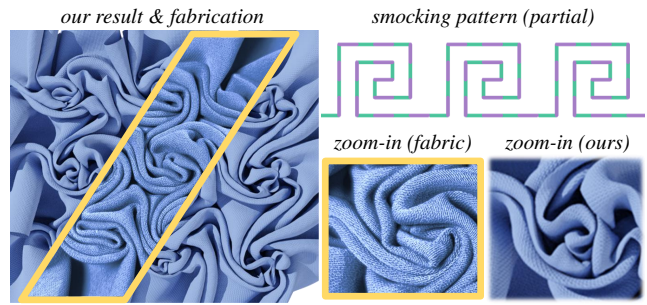

\begin{figure}[!t]
    \centering
    \includegraphics[width=\linewidth]{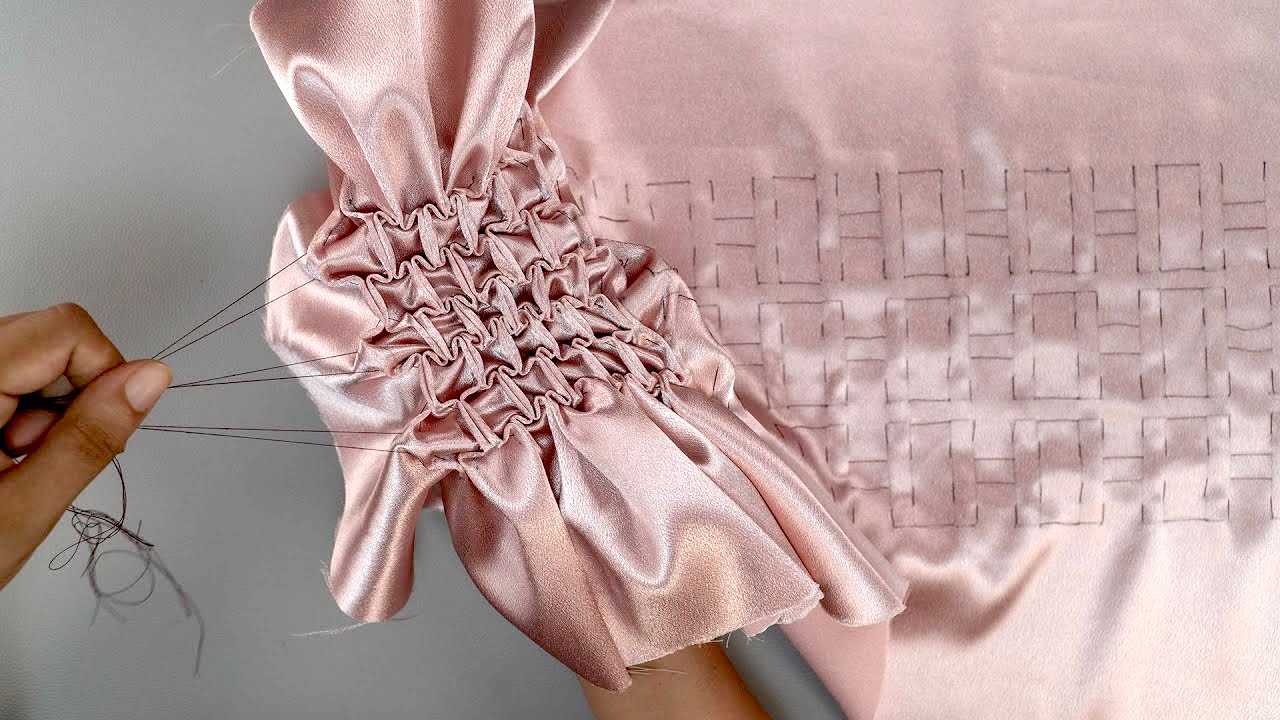}\vspace{-6pt}
    \caption{\textbf{Italian smocking fabrication}~\cite{egfabrication}. A single thread is sewn through the fabric along one row of the pattern. Then, the free ends of the threads are pulled on, resulting in intricate pleat patterns. \copyright HandiWorks YouTube channel. Used with permission.}
    \label{fig:intro:fabrication-process}\vspace{-6pt}
\end{figure}

In Italian smocking, the fabric is ``drawn up into close pleats on \emph{rows} of gauged stitching with deviations that create \emph{patterned irregularities}''~\cite{wolff1996art}.
% See Fig.~\ref{fig:intro:canadian-italian} (right) for an example pattern of Italian smocking.
% fabrication
An Italian smocking pattern is typically divided into rows (thread paths, not necessarily straight); each row is stitched with a separate thread. The thread alternates between passing in and out of the fabric, creating front stitches and back stitches.
After completing the stitching of all rows, the threads are gently pulled on, so that the fabric is pushed and gathered, creating folded pleats.
Finally, the thread ends are tied to secure the pleats in the desired shape.
See \figref{fig:teaser} for an example and 
\figref{fig:intro:fabrication-process} for a demonstration of the fabrication process.

\begin{figure}[!t]
    \centering
    \includegraphics[width=\linewidth]{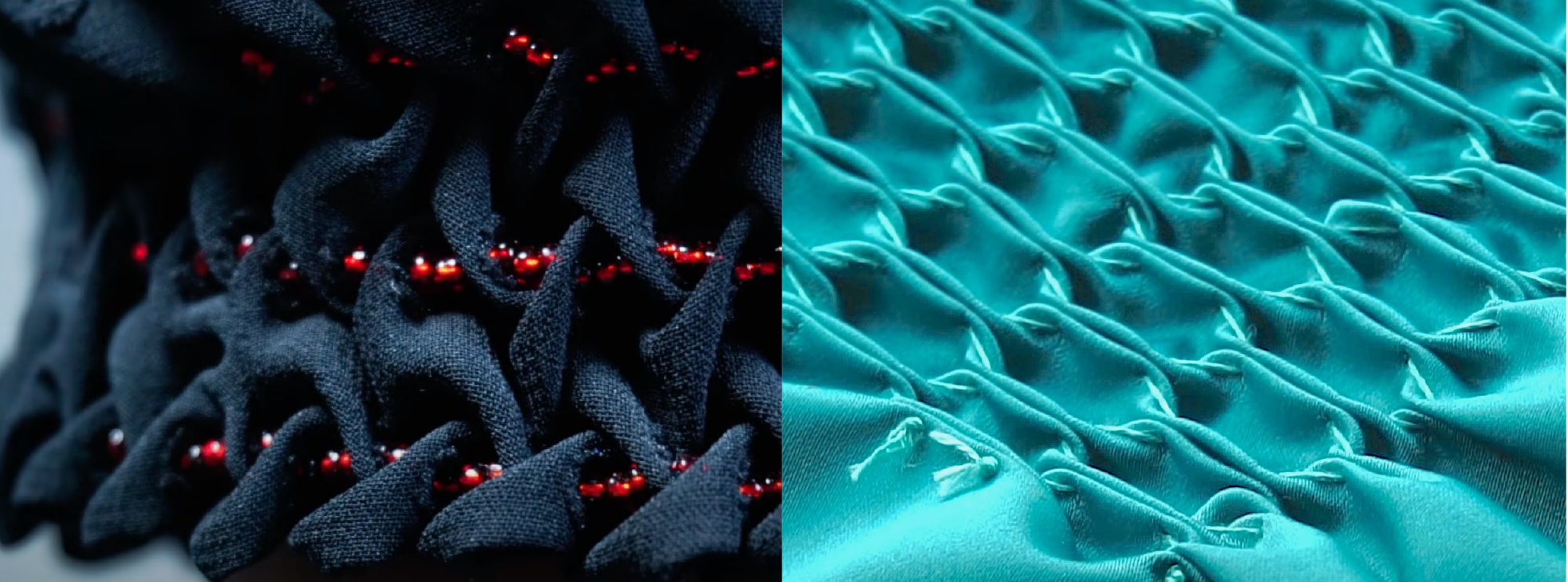}\vspace{-2pt}
    \caption{\textbf{Loose threads}. 
    An essential characteristic of Italian smocking is the deliberate avoidance of fully tightening the threads to achieve visually appealing pleats. \textbf{Left}: Beads are strategically employed to prevent the threads from being pulled taut~\cite{egbead}. \textbf{Right}: Visible threads provide structural support to the pleats and enhance the overall aesthetics in the final result~\cite{eghoney}. \copyright DIYstitching YouTube channel. Used with permission.}
    \label{fig:intro:fabrication-loose-thread}
\end{figure}

\begin{figure}[!t]
    \centering
    \begin{overpic}[trim=0cm 0cm 0cm -1cm,clip,width=1\linewidth,grid=false]{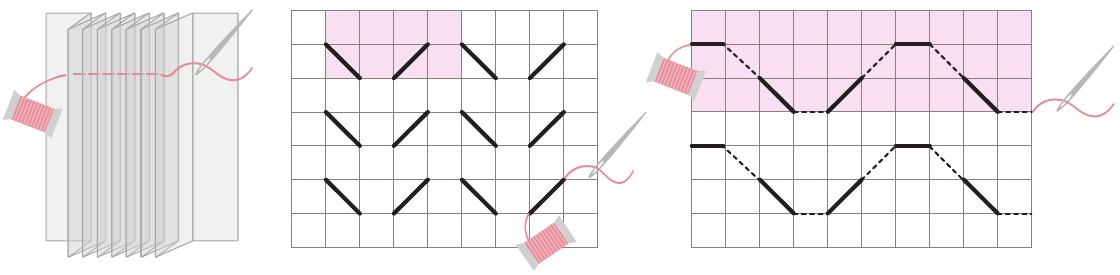}
    % \put(3,-3){\tiny English smocking}
    % \put(29,-3){\tiny Canadian smocking}
    % \put(68,-3){\tiny Italian smocking}
    \put(0,26){\footnotesize English smocking}    
    \put(26.5,26){\footnotesize Canadian smocking}    
    \put(66,26){\footnotesize Italian smocking}
    % \put(6,28){\footnotesize English}
    % \put(5,25){\footnotesize smocking}    
    % \put(34,28){\footnotesize Canadian}
    % \put(34,25){\footnotesize smocking}    
    % \put(72,28){\footnotesize Italian}
    % \put(70,25){\footnotesize smocking}
    \end{overpic}\vspace{-12pt}
    \caption{\new{
    Different types of smocking}. For the Italian smocking pattern, the front and back stitches are delineated in solid and dashed line segments, respectively.}
    \label{fig:intro:canadian-italian}
\end{figure}

% main diffrence to canadian smocking and the challenge
We can readily observe the distinctive characteristics of Italian smocking, setting it apart from English and Canadian smocking.
In English smocking, the fabric is stitched in straight paths with regular spacing, yielding uniform pleats.
The colorful rows of stitches running through these pleats are similar to standard 2D embroidery.
In contrast, the stitching lines in Italian smocking do not follow a straight path. As the stitching line turns and curves, the fabric gathers and crinkles along the stitched thread, resulting in complex and unpredictable pleats with diverse, irregular structures, which play the main decorative role.
Unlike the localized knots in Canadian smocking, which immobilize the stitching points and validate the assumption of negligible fabric thickness, Italian smocking permits the stitching points (i.e., the fabric) to move freely along the stitched thread, and the thread may not necessarily be pulled taut in the final fabrication (see \figref{fig:intro:fabrication-loose-thread} for some examples). We summarize the major differences between English, Canadian, and Italian smocking techniques in Table~\ref{tab:intro:smocking-types}\new{, and provide an illustrative comparison in \figref{fig:intro:canadian-italian}}.

\begin{table}[!b]
\caption{We compare three different types of smocking including English smocking, Canadian smocking, and Italian smocking.}
\label{tab:intro:smocking-types}
\centering
\definecolor{tabeasy}{HTML}{99d98c}
\definecolor{tabhard}{HTML}{f08080}
\definecolor{tabmed}{HTML}{ffd000}
\newcommand{\yes}{\cellcolor{mygreen!30}\cmark}
\newcommand{\medium}{\cellcolor{tabmed!20}\checked}
\newcommand{\no}{\cellcolor{myorange!30}\xmark}
\footnotesize
{\def\arraystretch{1.2}
\begin{tabular}{lccc}
\toprule[1pt]
\multicolumn{1}{l}{\textit{properties} \textbackslash \ smocking type} & English & Canadian & Italian \\ \midrule[1pt]
\textit{\textbf{continuous, long} stitching paths?} & \yes  & \no  & \yes \\
\textit{\textbf{localized} stitches and knots?}       & \no  & \yes  & \no \\ 
\textit{\textbf{regular} spacing?}                & \yes  & \no  & \no \\
\textit{stitching paths \textbf{pulled taut}?}    & \yes  & \yes  & \no \\
\textit{decoration from \textbf{colorful stitches}?} & \yes  & \no  & \yes \\
\textit{decoration from \textbf{voluminous pleats}?} & \no  & \yes  & \yes \\
\bottomrule[1pt]
\end{tabular}
}
\end{table}

In this work, we propose a novel method for previewing the result of Italian smocking given a stitching pattern by employing the general perspective of fabric-thread interaction.
We model the fabric using a coarse-grained mass-spring system at a scale similar to the size of the front and back stitches on the fabric, as indicated in the input pattern. 
Subsequently, we simulate the dynamics of this coarse mass-spring system by incorporating adjustable stitching springs to model taut threads during the thread-pulling stage of Italian smocking fabrication.
The resulting configuration of the mass-spring system is then used to guide the deformation of the fabric in a finer resolution by adapting {\cipc}~\cite{li2021CIPC}.

\emph{Contributions.}
(1) We introduce a formalization of Italian smocking pattern design and propose a simple method to preview the smocked results. 
(2) Our formulation is capable of handling continuous stitches on both the front and the back of the fabric, which can be used to facilitate free-form embroidery design.
%also presents an innovative approach to Canadian smocking, which typically involves isolated back stitches.
(3) We integrate non-zero sewing lengths and positional constraints into \new{the state-of-the-art simulator} {\cipc}~\cite{li2021CIPC}.

\section{Related work}\label{sec:related work}
Fabric manipulation, such as embroidery, folding, and pleating, is a fascinating art form that poses challenging problems in computer graphics.
These challenges include how to model the geometry of folds and how to efficiently and accurately simulate wrinkles. We therefore review the literature in these two aspects.

\emph{Modeling folds and pleats.}
Curves are popular for abstracting cloth folds.
For example,
Popa et al.~\cite{popa2009wrinkling} extract fold curves from input images to guide the 3D fabric deformation, achieving folds that closely match the reference. 
Jung et al.~\cite{jung2015sketching} propose a method to generate folds for garments or fashion accessories that are guaranteed to align with the curves drawn as silhouettes on a design sketch.
The FoldSketch system \cite{Li:2018:FoldSketch} solves for an updated sewing pattern that aligns with user-specified curves, achieving expected folds through a combination of graph editing and cloth simulation.
Zheng et al.~\cite{zheng2020foldGen} develop a system that allows users to generate folds on draped cloth by drawing curves and reconstructing the modified cloth surface for downstream simulation.
Frequency-based models are investigated to simulate the fine-scale wrinkles and their evolution on coarse meshes~\cite{chen2021fine, chen2023wrinkle}. 
Hermite radial basis functions can also be used to approximate the shape of wrinkles~\cite{mouhou2021wrinkle}.
However, while curves and smooth basis functions have proven effective in modeling prominent folds, they are less suitable for capturing the intricate and structural folds found in embroidery, especially at a fine level of detail.
In the special case of Canadian smocking, Kim~\cite{kim2020study} accounts for the fold angle when performing step-by-step simulation manually set up using the Clo software~\cite{CLO3D}.
Ren et al.~\cite{ren2023smocking} propose using a coarse graph to abstract the pleat shape and solve for the fine-grained shape of the pleats via surface deformation guided by this coarse graph.
However, none of the existing formulations can be applied to Italian smocking~\cite{wolff1996art,Naoko2013craft}, where the intricate pleats are formed by the special fabric-thread interaction.

\emph{Cloth simulation.}
Various elastic models~\cite{terzopoulos1987elastically} have been explored to simulate cloth dynamics, such as finite element representation~\cite{baraff1998large, narain2012adaptive}, mass-spring system \cite{choi2002cloth, Liu2013FSM}, and yarn-level model~\cite{kaldor2010efficient, cirio2014yarn}. Advanced general-purpose simulators can be adapted to use in cloth simulation~\cite{li2020incremental,li2021CIPC}.
The codimensional incremental potential contact model ({\cipc})~\cite{li2021CIPC} is the state-of-the-art simulator designed for precision, efficiency, and stability, built upon~\cite{li2020incremental}. {\cipc} achieves intersection-free and strain-limit-satisfied simulation through the introduction of additional barrier models for strain-limiting potentials, thickness boundaries, and continuous collision detection.
\new{Coarse-to-fine approaches are introduced to capture the details of simulated cloth. Bergou et at.~\cite{bergou2007tracks} augment coarse input with physically simulated details.}
% \new{TODO: add reference to use coarse proxy models to guide fine-level deformation.}
Zhang et al.~\cite{Zhang22progressive} propose a progressive method to accelerate cloth simulation for detailed wrinkles.
Existing physical simulators can produce plausible results without self-intersections for large-scale fabric manipulation, such as cloth draping.
However, when applied to fine-scale embroidery like smocking without geometric priors or precise guidance, simulators often struggle to create regular and realistic pleats~\cite{ren2023smocking}. 
These challenges are likely due to factors such as excessive contraction under the assumption of zero sewing length, asymmetric motion solvers, and inappropriate stitched thread representation.

\section{Preliminaries}\label{sec:background}

An Italian smocking pattern $\P$ is usually drawn on a piece of fabric with a set of \emph{stitching paths} $\L = \{\ell\}$. \new{Each stitching path is a list of \emph{consecutive} grid vertices with alternating front stitches and back stitches.}
\figref{fig:intro:canadian-italian} (right) shows an example of Italian smocking, which differs fundamentally from Canadian smocking in several ways:
(1) Canadian smocking patterns consist of \emph{disconnected} and short stitching lines, which are only annotated on \emph{one side} of the fabric. In contrast, Italian smocking patterns feature \emph{continuous} and long stitching paths that traverse both the \emph{front and back} sides of the fabric alternately.
(2) Creating a Canadian smocking pattern requires independently sewing each stitching line with localized knots. Conversely, Italian smocking patterns are fabricated along the stitching paths in a row-by-row fashion without localized knots.
(3) The decorative pleats are formed differently as well. In Canadian smocking, the localized stitches exert force to push the fabric out of the initial fabric plane, resulting in voluminous pleats. Consequently, the fabric's thickness can be largely ignored, assuming that after stitching, several points merge into one~\cite{ren2023smocking}. 
In contrast, for Italian smocking, the fabric is gently compressed and folded, and the resulting pleats take shape as the thread is pulled at the free ends. 
\new{These threads interweave a long sequence of points in the fabric and gather pleats. The thickness of the fabric plays a crucial role in maintaining the order of the created pleats and cannot be disregarded in physical simulation.}
% The thickness of the fabric plays a crucial role in the pleat formation process and cannot be disregarded. \new{TODO: add a little more explanation of why the thickness matters to Italian smocking.}

We model an input pattern $\P$ by a mass-spring system $(\V, \E)$, where $\V$ is the set of vertices from the original grid and $\E$ is the set of edges (springs) that connect the vertices in $\V$.
Specifically, there are two types of springs in $\E$: (1) the \textbf{fabric springs} $\Ef$ that connect the adjacent grid vertices in $\V$, and (2) the \textbf{stitching springs} $\Es$ that connect two stitching vertices of a line in $\L$. 
%We denote $\Es\left(\ell\right)$ the subset of stitching springs that correspond to the stitching path $\ell\in\L$.
%We can readily observe that the arrangement of the stitching springs $\Es$ is consistent with the stitching path annotation $\L$, and $\Es = \bigcup_{\ell\in\L} \Es\left(\ell\right)$.

\begin{figure}[!t]
    \centering
    \includegraphics[width=1\linewidth]{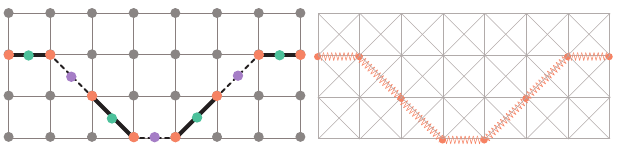}
    \caption{\textbf{Mass-spring system}. For the Italian smocking pattern shown on the left, with front (resp.\ back) stitches annotated in solid (resp.\ dashed) line segments, we define a mass-spring system based on the original grid. \textbf{Left}:  we highlight the stitching vertices $\Vs$, front midpoints $\Vf$, back midpoints $\Vb$, and pleat vertices $\Vp$, in \textcolor{myorange}{orange}, \textcolor{mygreen}{green}, \textcolor{mypurple}{purple}, and \textcolor{gray}{gray}, respectively. 
    \textbf{Right}: we color the fabric spring (stitching spring) in \textcolor{gray}{gray} (\textcolor{myorange}{orange}).}\label{fig:background:notation}
\end{figure}

We categorize the vertices in $\V$ into two groups, see also \figref{fig:background:notation}: (1) $\Vs$: the \textbf{stitching vertices}  that lie on the stitching paths $\L$, and (2) $\Vp = \V \backslash \Vs$: the remaining vertices, which we refer to as \textbf{pleat vertices}.
Note that the encoding of the front-back information is not possible within the stitching springs because there are no faces to distinguish the front or back side of a fabric in the mass-spring system. 
We therefore sample the midpoints of all front and back stitches in the stitching paths to encode the front-back information.
Specifically, $\Vf$ denotes the midpoints of the \emph{front stitches} in $\L$ (we call them \textbf{front midpoints} for simplicity) and $\Vb$ denotes the midpoints of the \emph{back stitches} in $\L$ (called \textbf{back midpoints}). See \figref{fig:background:notation} (left) for an illustration.
For a vertex $v_i \in \V\cup\Vf\cup\Vb$, we denote its original 2D position in the flat fabric as $\overline{\matr{x}}_i\in\R^2$.

% \subsection{Canadian smocking vs.\ Italian smocking}
% Canadian smocking is a popular embroidery technique that creates intricate and voluminous pleats from localized stitches. Fig.~\ref{fig:intro:canadian-italian} (left) shows an example Canadian smocking pattern, where during fabrication, the endpoints of each stitching line (delineated by black line segments) are gathered and sewn together. 
% In practice, a Canadian smocking pattern is generated by evenly repeating a unit pattern, highlighted by the pink region, across the fabric. 
% Ren et al.~\cite{ren2023smocking} propose an efficient method to preview the smocked design as a surface in 3D, computed from an input 2D Canadian smocking pattern.
% Specifically, a smocked graph is extracted from the input pattern by fusing all stitching nodes that belong to the same stitching line into a single node and eliminating degenerated or duplicated edges.
% This fusion process results in a non-manifold smocked graph, which constitutes a coarse representation of the fabric 
% %altering the geodesic distances among the nodes, 
% and is subsequently embedded in 3D.
% Finally, the embedded smocked graph is used to guide the deformation of the fabric in a much finer resolution.
% %via the as-rigid-as-possible ({\arap}) method~\cite{arap}.

\section{Method}\label{sec:mtd}

% \begin{figure*}[!t]
%     \centering
%     \begin{overpic}[trim=0cm 0cm 0cm -0.5cm,clip,width=1\linewidth,grid=false]{figures/eg_curve_shrinkage_iter.pdf}
%     \put(1.8,16.5){\footnotesize (a) smocking pattern}
%     \put(27,16.5){\footnotesize (b) our results with different shrinkage}
%     \put(97,2.5){\footnotesize (c)}
%     \put(24,0){\footnotesize $\gamma = 50\%$}
%     \put(40,0){\footnotesize $\gamma = 30\%$}
%     \put(51.8,0){\footnotesize $\gamma = 10\%$}
%     \put(68,0){\footnotesize front}
%     \put(88,0){\footnotesize back}
%     \end{overpic}
%     \caption{For the \textsc{Curve} pattern shown in (a) where the front (resp.\ back) stitches are colored in \textcolor{mygreen}{green} (resp.\ \textcolor{mypurple}{purple}), we run our algorithm using various values of the shrinkage parameter $\gamma$, as illustrated in three examples in (b). In (c), we show the front and back sides of a zoomed-in version of our result at a shrinkage level of $\gamma=10\%$.}
%     \label{fig:mtd:curve-diff-shrinkage}
% \end{figure*}

Given an input Italian smocking pattern, our goal is to preview the corresponding fabricated result in the form of a surface mesh in 3D, showing the intricate folds and pleats.
The main challenges of this problem include accounting for the fabric thickness, distinguishing between front and back stitches, and determining the non-vanishing length of stitched threads in the fabricated state.
%: (1) The thickness of the fabric cannot be neglected since we need to distinguish the front stitches from the back stitches. (2) A stitching spring is not necessarily in zero-length after fabrication. 
%Instead, since there is only a single thread following one row of the pattern and sewing through the fabric several times without localized knots, the stitching vertices have considerable freedom to move along the thread. This makes it hard to pose any assumptions on the expected length of each stitching spring.
%
\new{Existing methods such as \cite{ren2023smocking} and {\cipc}~\cite{li2021CIPC} do not support advanced stitching primitives. If the smocking stitches are considered as seams, these methods struggle with Italian smocking, since the distance between two stitching points is assumed to be zero.} 
% Existing methods, including \cite{ren2023smocking} and {\cipc}~\cite{li2021CIPC}, struggle with Italian smocking, since they assume the stitching springs to be zero-length. 
% \new{TODO: add more explanation of the configuration of simulator. Our method only address the limitation of these specific simulators other than other stitching primitives with practical benefits.}

To tackle the challenges, we propose an abstraction of a mass-spring system with front-back information from the input pattern. We then dynamically estimate the expected lengths of the stitching springs and find the 2D embedding of the stitching vertices via the simulated configuration of the mass-spring system (\secref{sec:mtd:2d}).
The resulting spring lengths and positions of the stitching vertices are fed into {\cipc} to guide the deformation of the fabric in fine resolution, with collision handling (\secref{sec:mtd:3d}).

\subsection{2D simulation of the mass-spring system}\label{sec:mtd:2d}
Drawing inspiration from~\cite{ren2023smocking}, we extract a low-resolution mass-spring system $\left(\V, \E\right)$ to distill the geometric priors from the input Italian smocking pattern.
Solving the dynamics of such a mass-spring system appears straightforward at first glance.
However, the fact that the expected lengths for all fabric springs $\Ef$ and the stitching springs $\Es$ are \emph{unknown} makes the problem hard to formulate and solve.
For example, as shown in \figref{fig:intro:fabrication-loose-thread}, the final distance between two consecutive stitching vertices is not necessarily zero, leaving the expected length of a stitching spring unpredictable.
Furthermore, in our setting, the length of a fabric spring in the final embedding is often much smaller than its original length in the fabric.
%Specifically, in typical cases where a high-resolution mass-spring system is employed to model a small fabric section, it is common to assume that fabric springs have the final-state length equal to their original length. This rest length indicates the anticipated Euclidean distance between two vertices following fabric deformation.
Specifically, in typical cases where a high-resolution mass-spring system is employed to model a small fabric piece, it is common to assume that fabric springs preserve their original lengths during fabric deformation. Here the length indicates the anticipated Euclidean distance between two vertices.
However, this assumption is not applicable to our low-resolution mass-spring system. 
For a fabric spring shown in \figref{fig:background:notation} (right),
with the thread tightened, its two endpoints would be closely embedded at a distance equal to the fabric thickness after smocking, which significantly deviates from the spring's original length within the fabric.

\begin{figure}[!t]
    \centering
    \begin{overpic}[trim=8cm 0.8cm 18.8cm 1cm,clip,width=1\linewidth,grid=false]{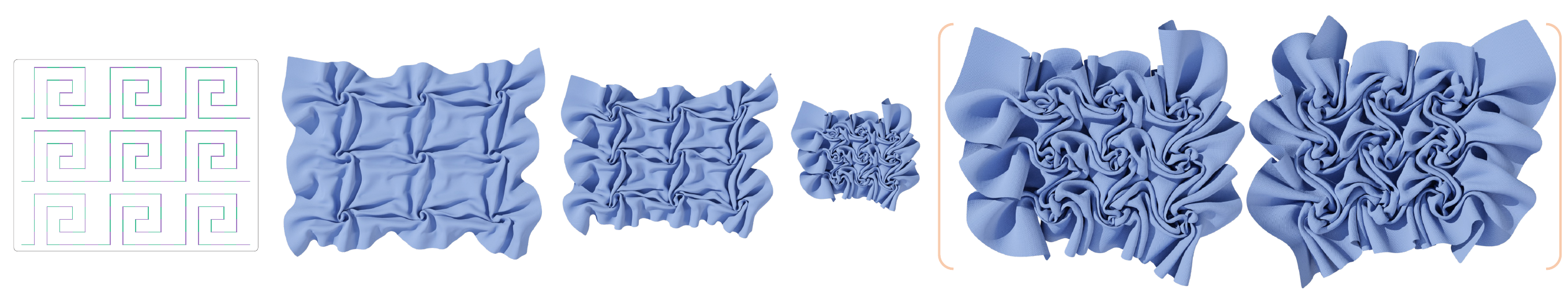}
    \put(16,32){\footnotesize $\gamma = 50\%$}
    \put(54,32){\footnotesize $\gamma = 30\%$}
    \put(82,32){\footnotesize $\gamma = 10\%$}
    \end{overpic}
    \caption{Our results with different shrinkage $\gamma$.}
    \label{fig:mtd:curve-diff-shrinkage}
\end{figure}

To address these challenges, we propose solving for a constrained \emph{2D projection} of the dynamics of the abstracted mass-spring system, rather than dealing with the full 3D mass-spring system with unknown expected lengths. 
More precisely, we would like to find a 2D projection $\matr{X}$ for the mass-spring system by solving:
\begin{subequations}\label{eq:mtd:2d:prob:all}
\begin{align}
\max_{\matr{X}\in\R^{\vert \V \vert \times 2}} & \quad \sum_{(i,j)\in \E} \left\Vert \matr{x}_i - \matr{x}_j \right\Vert_2, \label{eq:mtd:2d:prob:energy}\\
\text{s.t.} & \quad \tau \le \left\Vert \matr{x}_p - \matr{x}_q \right\Vert_2 \le  \left\Vert \overline{\matr{x}}_p - \overline{\matr{x}}_q \right\Vert_2, \ \ \ \forall (p,q)\in\Ef ,\label{eq:mtd:2d:prob:fabric-spring}\\
&  \quad \tau \,\, \vert \Es \vert  \le \D\left(\matr{X} \mid \L \right) \le \gamma\,\, \D\left(\overline{\matr{X}} \mid \L \right), \label{eq:mtd:2d:prob:stitching-spring}
\end{align}
\end{subequations}
where $\tau$ is the fabric thickness, $\vert\Es\vert$ is the number of edges (springs) in the stitching paths $\L$, $\overline{\matr{x}}_i$ is the original 2D position of the vertex $v_i$ and $\overline{\matr{X}}$ is the set of all these original positions. 
The function $\D(\matr{X}\mid\L)$ measures the total length of the threads that pass through the fabric along the stitching paths $\L$ given the current embedding  $\matr{X}$:
\begin{equation}\label{eq:mtd:2d:thread-length}
    \D\left(\matr{X} \mid \L\right) = \sum_{(i, j)\in \E_s} \Vert \matr{x}_i - \matr{x}_j\Vert_2.
\end{equation}

The energy in \eqnref{eq:mtd:2d:prob:energy} encourages the vertices to stay away from each other to avoid cluttered pleats, similar to~\cite{ren2023smocking}.
The first set of constraints, \eqnref{eq:mtd:2d:prob:fabric-spring}, poses restriction on the length of a fabric spring $(p, q)\in \Ef$: once embedded in 2D, 
it must be at minimum the fabric thickness $\tau$ and at maximum equal to the original spring length of the initial flat fabric state. A violation of these constraints can result in fabric penetration or tearing.

The second set of constraints, \eqnref{eq:mtd:2d:prob:stitching-spring}, poses restrictions on the stitching springs. 
Recall that when fabricating an Italian smocking pattern, the free ends of the threads (aligned with the stitching paths $\L$) are pulled. This action causes the fabric to fold and pleat along the stitching paths until the desired texture is achieved. 
The in-between stitching vertices have the freedom to slide along the thread. 
We therefore consider the total length of the threads $\D(\matr{X}\mid\L)$ instead of each stitching spring independently.
The minimal possible total length is $\tau\,\vert\Es\vert$, which corresponds to the scenario when the threads are pulled completely taut, such that no stitching vertices can slide.
At the same time, we propose a natural upper bound where the sum of the thread lengths is reduced to a $\gamma$ fraction of their original total length.
Here, the hyper-parameter $\gamma$ models the pulling force applied to the threads. 
For example, when $\gamma=1$, we can see that $\matr{X} = \overline{\matr{X}}$ gives the optimal solution. A smaller value $\gamma < 1$ drives the stitching vertices to move closer to meet the constraints, leading to folds and pleats. See \figref{fig:mtd:curve-diff-shrinkage} for an example.

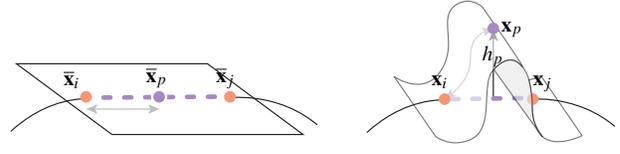
\begin{figure}[!t]
    \centering
    \begin{overpic}[trim=0cm 0cm 0cm 0cm,clip,width=1\linewidth,grid=false]{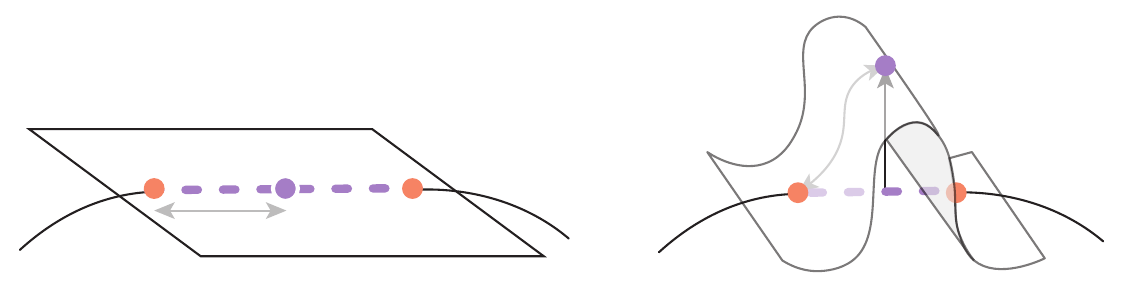}
    \put(10,11.5){\footnotesize $\overline{\matr{x}}_i$}
    \put(34,12){\footnotesize $\overline{\matr{x}}_j$}
    \put(23,12){\footnotesize $\overline{\matr{x}}_p$}
    % \put(18,4.3){\footnotesize $\overline{d}$}

    \put(68,11.5){\footnotesize $\matr{x}_i$}
    \put(84,11.5){\footnotesize $\matr{x}_j$}
    \put(79,20){\footnotesize $\matr{x}_p$}
    \put(76,15.4){\footnotesize $h_p$}
    \end{overpic}
    \caption{Pulling the thread of a back stitch (left) causes the fabric to bend outward (right). We can estimate the height $h_p$ of the midpoint $\matr{x}_p$ based on Pythagoras' theorem, as shown in Eq.~\eqref{eq:mtd:height}.}
    \label{fig:mtd:height}
\end{figure}

\emph{Algorithm overview.} 
The non-linearly constrained, non-convex problem in \eqnref{eq:mtd:2d:prob:all} is hard to optimize directly. 
We therefore propose to solve for the embedding $\matr{X}$ in an alternating scheme: 
\begin{enumerate}[label={\bfseries{Step \arabic*}:},leftmargin=1cm]
\item initialize the embedding $\matr{X}^{(0)} = \overline{\matr{X}}$; 
\item estimate the expected distances $d_{ij}^{(k)}$ from the constraints based on $\matr{X}^{(k)}$ (detailed below);
% \item update the embedding $\matr{X}^{(k+1)}$ to match the expected distances $d_{ij}^{(k)}$ as an unconstrained problem and apply explicit Euler method for one timestep:
% \new{TODO: revise "Explicit Euler"}
\item update the embedding $\matr{X}^{(k+1)}$ \new{based on the expected distances $d_{ij}^{(k)}$, which is solved as an unconstrained problem:}
% \begin{equation}\label{eq:mtd:2d:mass-spring}
%     \matr{X}^{(k+1)} = \matr{X}^{(k)} - \delta_t\nabla_{\matr{X}} \sum_{(i, j)\in \E}\left(\left\Vert \matr{x}_i - \matr{x}_j\right\Vert_2 - d_{ij}^{(k)} \right)^2;
% \end{equation}
\begin{equation}\label{eq:mtd:2d:mass-spring}
    \matr{X}^{(k+1)} = \argmin_{\matr{X}\in\R^{\vert\V\vert\times 2}}\quad \sum_{(i, j)\in \E}\left(\left\Vert \matr{x}_i - \matr{x}_j\right\Vert_2 - d_{ij}^{(k)} \right)^2;
\end{equation}
\item set $k\leftarrow k+1$ and go to Step 2, unless a stopping criterion is satisfied.
\end{enumerate}

We now detail the estimation of $d_{ij}^{(k)}$ based on the current embedding $\matr{X}^{(k)}$ mentioned in Step 2.
For a fabric spring $(i,j)\in \Ef$, \new{$d_{ij}^{(k)}$ is determined to make sure the constraints in Eq.~\eqref{eq:mtd:2d:prob:fabric-spring} are satisfied:}
\begin{equation}\label{eq:mtd:2d:dij:fabric}
 d_{ij}^{(k)} = \max \left\{\tau,\ \min\left\{\Vert \matr{x}_i^{(k)} - \matr{x}_j^{(k)} \Vert_2,  \Vert \overline{\matr{x}}_i - \overline{\matr{x}}_j\Vert_2\right\}\right\}. 
\end{equation} 
\new{
For stitching springs, we introduce a pulling direction to model their shrinkage speed. In physical fabrications, we observe that the stitches that align more closely with the pulling direction (i.e., with a larger projected length) shrink more rapidly (i.e., have a smaller expected length $d_{ij}^{(k)}$), and vice versa. Specifically,}
for a stitching spring $(i,j)\in\Es\left(\ell\right)$, we compute 
\begin{equation}\label{eq:mtd:2d:pull-direction}
d_{ij}^{(k)} = \max \left\{ \tau, \ \proj_{\matr{d}}\left( \matr{x}_i^{(k)} - \matr{x}_j^{(k)}\right) \right\}, 
\end{equation}
where $\proj_{\matr{d}}\left( \matr{x}\right)$ gives the projected length of a vector $\matr{x}$  onto the direction $\matr{d}$, which is orthogonal to the pulling direction. 
\new{Here, for notation simplicity, instead of expressing the expected length as inversely proportional to the projection onto the pulling direction, we represent it as proportional to the projection onto the orthogonal of the pulling direction.}
% \new{TODO: explain more for the pulling direction. Why use pulling direction \& why orthogonal. }
The upper bound in \eqnref{eq:mtd:2d:prob:stitching-spring} is adopted as the stopping criterion in Step 4. Specifically, we stop the algorithm when the total length of the threads is below the preset shrinkage ratio $\gamma$. \new{See Algorithm~\ref{algo:2d-sim} for a detailed description and Appendix~\ref{app:convergence} for its convergence behavior analysis.
In Appendix~\ref{app:sqp}, we provide a comparison to an existing solver, sequential least squares programming (SLSQP)~\cite{kraft1988slsqp}, where our algorithm shows significantly faster convergence and leads to better simulated results.
}

\begin{algorithm}[!b]
\caption{\new{2D simulation of the mass-spring system}} \label{algo:2d-sim}
\textbf{Hyper-parameters:} shrinkage $\gamma$, projection direction $\matr{d}$, fabric thickness $\tau$, time step $\Delta t =\unit[0.1]{sec}$, damping ratio $\alpha = 0.9$, fabric spring stiffness $k_f = 1$, stitching spring stiffness $k_s = 5$.
\begin{algorithmic}[1]
\Procedure{2D Simulation}{$\matr{X},\overline{\matr{X}},\L, \Es,\Ef,\E$}
    \State $\matr{a} \leftarrow \matr{0}, \matr{v} \leftarrow \matr{0}$

    \While{$\D\left(\matr{X} \mid \L \right) > \gamma\,\, \D\left(\overline{\matr{X}} \mid \L\right)$}  \Comment{{\footnotesize{Stopping criterion}}}
        
        \For{$(i,j)\in \Es$}     \Comment{\footnotesize{Expected length for stitching springs}}
            \State $d_{ij} \leftarrow \max \left\{ \tau, \ \proj_{\matr{d}}\left( \matr{x}_i - \matr{x}_j\right) \right\}$
            \Comment{\footnotesize{\eqnref{eq:mtd:2d:dij:fabric}}}
            \EndFor
            
        \For{$(i,j)\in \Ef$}     \Comment{\footnotesize{Expected length for fabric springs}}
            \State $d_{ij} \leftarrow \max \left\{\tau,\ \min\left\{\Vert \matr{x}_i - \matr{x}_j \Vert_2,  \Vert \overline{\matr{x}}_i - \overline{\matr{x}}_j\Vert_2\right\}\right\}$
            \Comment{\footnotesize{\eqnref{eq:mtd:2d:pull-direction}}}
            \EndFor

        \For{$(i,j)\in \E$}      \Comment{\footnotesize{Calculate acceleration}}
            \State $\Delta \matr{d}_{ij} \leftarrow (\Vert \matr{x}_i - \matr{x}_j\Vert_2 - d_{ij}) \nicefrac{ \left(\matr{x}_i - \matr{x}_j\right)}{\Vert \matr{x}_i - \matr{x}_j\Vert_2}$
            \State $\mathbf{a}_i \leftarrow \mathbf{a}_i - k \Delta \matr{d}_{ij}$ 
            \State $\mathbf{a}_j \leftarrow \mathbf{a}_j + k \Delta \matr{d}_{ij}$ 
            
            \EndFor

        \State $\mathbf{v} \leftarrow \mathbf{v} + \mathbf{a}\Delta t$
        \State $\mathbf{v} \leftarrow \alpha\mathbf{v}$
        \State $\mathbf{x} \leftarrow \mathbf{x} + \mathbf{v}\Delta t$
        \State $\matr{a} \leftarrow \matr{0}$   \Comment{\footnotesize{Reset for next timestep}}
    \EndWhile 

\EndProcedure

\end{algorithmic}
\end{algorithm}

\subsection{Mesh deformer via {\cipc}}\label{sec:mtd:3d}
\new{We use the solved 2D embedding $\matr{X}_s$ of the \emph{stitching vertices} to guide the 3D deformation of the fabric, which is now represented in a much finer resolution.
During the 2D simulation, the fabric springs $\Ef$ mainly serve as the inextensible constraints for the stitching springs $\Es$, and their actual positions, depending on the final shape of the formed pleats, are undetermined at this stage. We therefore only use the stitching points to guide the deformation, since they are better constrained to provide more accurate control.}
% \new{It is worth noting that the fabric springs $\Ef$ mainly serve as the inextensible constraints for the stitching springs $\Es$. Their actual positions depend on the final shape of the voluminous pleats, which are undetermined in the 2D simulation. Thus, only the fully constrained stitching vertices $\Vs$ are extracted to guide further deformation.}
% \new{Therefore, the overlapped fabric springs around the corner of the 2D results will not induce intersection in 3D deformation, as shown in \textbf{Fig.XXX}.}

\new{Specifically,} we first determine the 3D positions of the stitching vertices by assuming they all share the same height and experience no unbalanced external forces. Without loss of generality, we set the height to zero, i.e., the estimated 3D position for the stitching vertex, denoted $\matr{x}_s^c\in \R^3$, is:
\begin{equation}
\matr{x}_s^c \leftarrow \left( \matr{x}_s,\, 0\right), \quad  v_s\in\Vs.
\end{equation}
Next, we estimate the 3D position of the front/back midpoint $v_m$ of the stitching spring $(i,j)\in\Es$:
\begin{equation}
    \R^3\ni\matr{x}_m^c \leftarrow
    \begin{cases} 
      \left( \nicefrac{(\matr{x}_i + \matr{x}_j)}{2},\, - h_m\right) & \text{if }\, v_m\in\Vf \\
      \left( \nicefrac{(\matr{x}_i + \matr{x}_j)}{2}, \quad h_m\right) & \text{if }\, v_m\in\Vb
   \end{cases}
\end{equation}
The superscript $c$ abbreviates ``constrained'', as these positions serve as positional constraints.
Intuitively, the front stitches cause midpoints to fold inward, resulting in a negative height ($-h_m$), while the back stitches cause midpoints to fold outward, resulting in a positive height.
A good approximation for $h_m$ is (See \figref{fig:mtd:height}):
\begin{equation}\label{eq:mtd:height}
    h_m^2 + \left(\nicefrac{\left\Vert \matr{x}_i - \matr{x}_j\right\Vert}{2}\right)^2 = \left(\nicefrac{\left\Vert \overline{\matr{x}}_i - \overline{\matr{x}}_j\right\Vert}{2}\right)^2.
\end{equation}
For simplicity one can also estimate $h_m$ as $\nicefrac{\left(\left\Vert\overline{\matr{x}}_i - \overline{\matr{x}}_j\right\Vert - \left\Vert{\matr{x}}_i - {\matr{x}}_j\right\Vert\right)}{2}$ when the fold is vertical with a flat crease.

We augment the original incremental potential of {\cipc} \cite{li2021CIPC} with the positional constraints of the embedded stitching vertices $\matr{X}_s$, the estimated midpoints $\matr{X}_m$, and the sewing length constraints to obtain the deformed fabric mesh in 3D:
% \begin{equation}
%  \min_{\matr{X}\in \R^{n\times 3}} E_{\text{cipc}}\left(\matr{X}\right) + w_s \Delta t^2 E_{\text{sew}}\left(\matr{X} \mid d\left(\matr{X}_u\right) \right) + w_p E_{\text{pos}} \left(\matr{X} \mid \matr{X}_u \cup\matr{X}_m\right)  
% \end{equation}
\begin{equation}\label{eq:mtd:cipc-energy}
 \min_{\matr{X}\in \R^{n\times 3}} E_{\text{cipc}}\left(\matr{X}\right) + w_s E_{\text{sew}}\left(\matr{X} \right) + w_p E_{\text{pos}} \left(\matr{X}\right), 
\end{equation}
where $n$ is the total number of vertices in the finer representation of the fabric. 
The coefficients $w_s$ and $w_p$ are weights for these constraints. The term $E_{\text{sew}}$ encourages the sewing length between two stitching vertices to be equal to our computed values.
We add $\Delta t^2$, the squared timestep of {\cipc}, to $E_{\text{sew}}$, incorporating it into the incremental potential as the sewing energy potential:
\begin{align}
E_{\text{sew}}\left(\matr{X} \right)   = 
\Delta t^2 \! \sum_{(p, q)\in \Es}\left(\left\Vert \matr{x}_p - \matr{x}_q\right\Vert_2 - \left\Vert \matr{x}^c_p - \matr{x}^c_q\right\Vert_2 \right)^2.
\end{align}
The term $E_{\text{pos}}$ is the positional regularizer that encourages the stitching vertices and midpoints to remain close to the solved positions $\matr{X}_s \cup\matr{X}_m$:
\begin{equation}
E_{\text{pos}}\left(\matr{X}\right)  = 
\sum_{i\in \Vs\cup\Vf\cup\Vb}
\left\Vert \matr{x}_i - \matr{x}_i^c\right\Vert_2^2.
\end{equation}
\new{
We consider $E_{\text{pos}}$ as a soft positional constraint instead of a virtual energy potential. Therefore the timestep $\Delta t^2$ is not included.} 
Integrating these two regularizers into {\cipc} effectively guides the fabric from its initial flat state toward the desired pleat shapes, while the original configurations of {\cipc} handle self-collisions during the cloth simulation.

% \begin{subequations}\label{eq:mtd:2d:prob:all}
% \begin{align}
% \max_{\matr{X}\in\R^{\vert \V \vert \times 2}} & \quad \sum_{(i,j)\in \E} \left\Vert \matr{x}_i - \matr{x}_j \right\Vert_2, \label{eq:mtd:2d:prob:energy}\\
% \text{s.t.} & \quad \tau \le \left\Vert \matr{x}_p - \matr{x}_q \right\Vert_2 \le  \left\Vert \overline{\matr{x}}_p - \overline{\matr{x}}_q \right\Vert_2, \ \ \ \forall (p,q)\in\Ef ,\label{eq:mtd:2d:prob:fabric-spring}\\
% &  \quad \tau \,\, \vert \Es\left(\ell\right) \vert  \le \D\left(\matr{X} \mid \ell \right) \le \gamma\,\, \D\left(\overline{\matr{X}} \mid \ell\right), \quad\forall \ell\in\L, \label{eq:mtd:2d:prob:stitching-spring}
% \end{align}
% \end{subequations}
% %
% where $\tau$ is the fabric thickness, $\vert\Es\left(\ell\right)\vert$ is the number of edges (springs) in stitching line $\ell$, $\overline{\matr{x}}_i$ is the original 2D position of the vertex $v_i$ and $\overline{\matr{X}}$ is the set of all these original positions. 
% The function $\D(\matr{X}\mid\ell)$ measures the total length of the thread that passes through the fabric along stitching path $\ell$ given the current embedding  $\matr{X}$:
% \begin{equation}\label{eq:mtd:2d:thread-length}
%     \D\left(\matr{X} \mid \ell\right) = \sum_{(i, j)\in \E_s\left(\ell\right)} \Vert \matr{x}_i - \matr{x}_j\Vert_2.
% \end{equation}

\begin{figure*}[!t]
    \centering
    \begin{overpic}[trim=0cm 0cm 0cm -1cm,clip,width=1\linewidth,grid=false]{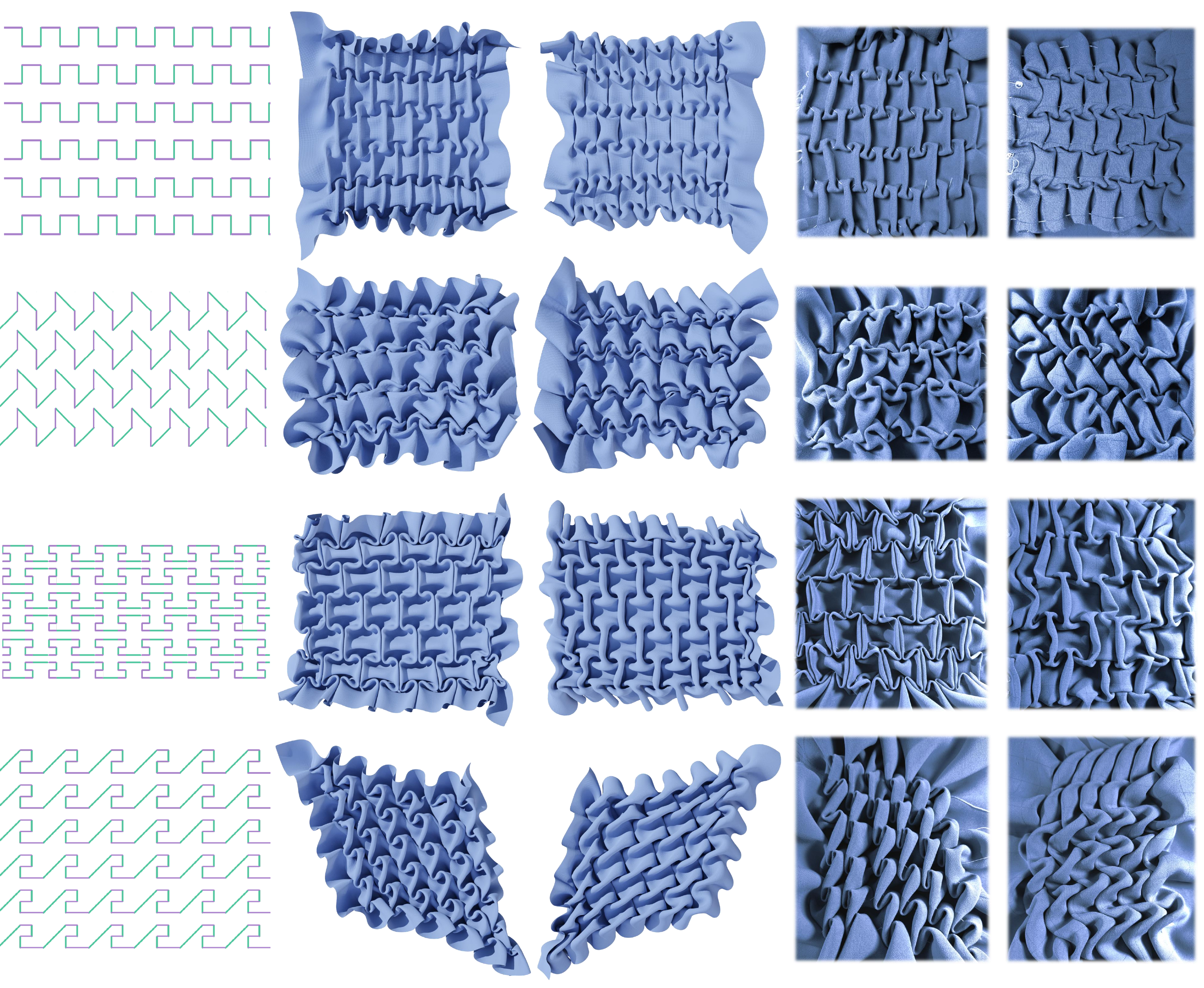}
    \put(-2,70){\footnotesize (a)}
    \put(-2,50){\footnotesize (b)}
    \put(-2,30){\footnotesize (c)}
    \put(-2,10){\footnotesize (d)}
    \put(41,80){\footnotesize $\gamma=30\%$}
    \put(41,60){\footnotesize $\gamma=20\%$}
    \put(41,41.5){\footnotesize $\gamma=20\%$}
    \put(41,18){\footnotesize $\gamma=20\%$}
    \put(5,81.5){\footnotesize smocking pattern}
    \put(30,81.5){\footnotesize ours (front)}
    \put(50,81.5){\footnotesize ours (back)}
    \put(68,81.5){\footnotesize fabrication (front)}
    \put(86,81.5){\footnotesize fabrication (back)}
    \end{overpic}\vspace{-3pt}
    \caption{We compare our simulated results to physical fabrications.}
    \label{fig:res:multi-row-patterns}
\end{figure*}

\begin{figure*}[!t]
    \begin{overpic}[trim=0cm 19cm 3cm -1cm,clip,width=1\linewidth,grid=false]{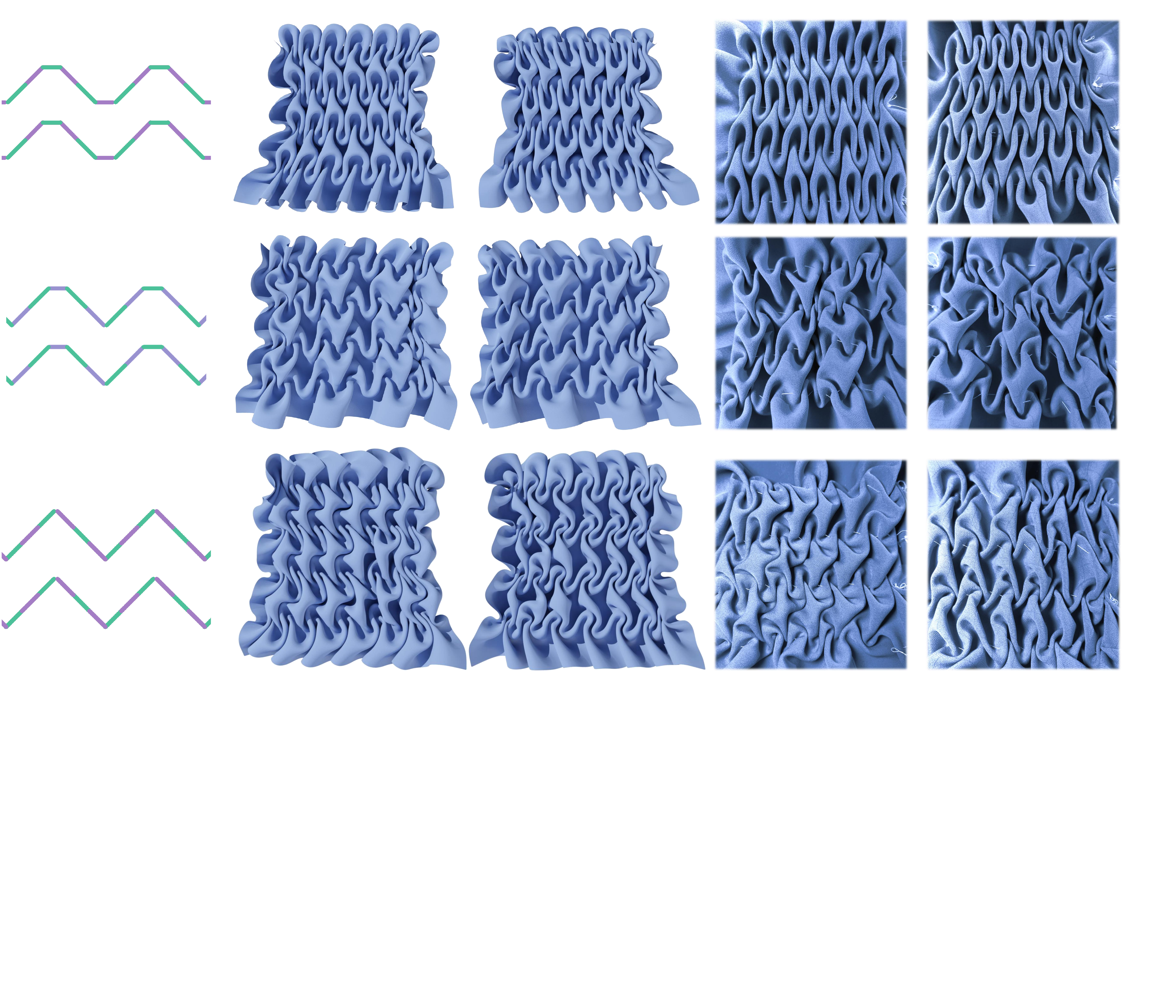}
    \put(-1,56){\footnotesize (a)}
    \put(-1,36){\footnotesize (b)}
    \put(-1,16){\footnotesize (c)}
    \put(7,45){\footnotesize $\gamma=20\%$}
    \put(7,25){\footnotesize $\gamma=30\%$}
    \put(7,4){\footnotesize $\gamma=20\%$}
    \put(4,60){\footnotesize smocking pattern}
    \put(6,58){\footnotesize (zoom-in)}
    \put(28,60){\footnotesize ours (front)} 
    \put(48,60){\footnotesize ours (back)}
    \put(66,60){\footnotesize fabrication (front)}
    \put(85,60){\footnotesize fabrication (back)}
    \end{overpic}\vspace{-9pt}
    \caption{Small changes in the smocking pattern can lead to significantly different pleats.
    }
    \label{fig:res:zigzag}
\end{figure*}

\begin{figure*}[!t]
    \centering
    \begin{overpic}[trim=0cm 1.5cm 0cm 0.5cm,clip,width=1\linewidth,grid=false]{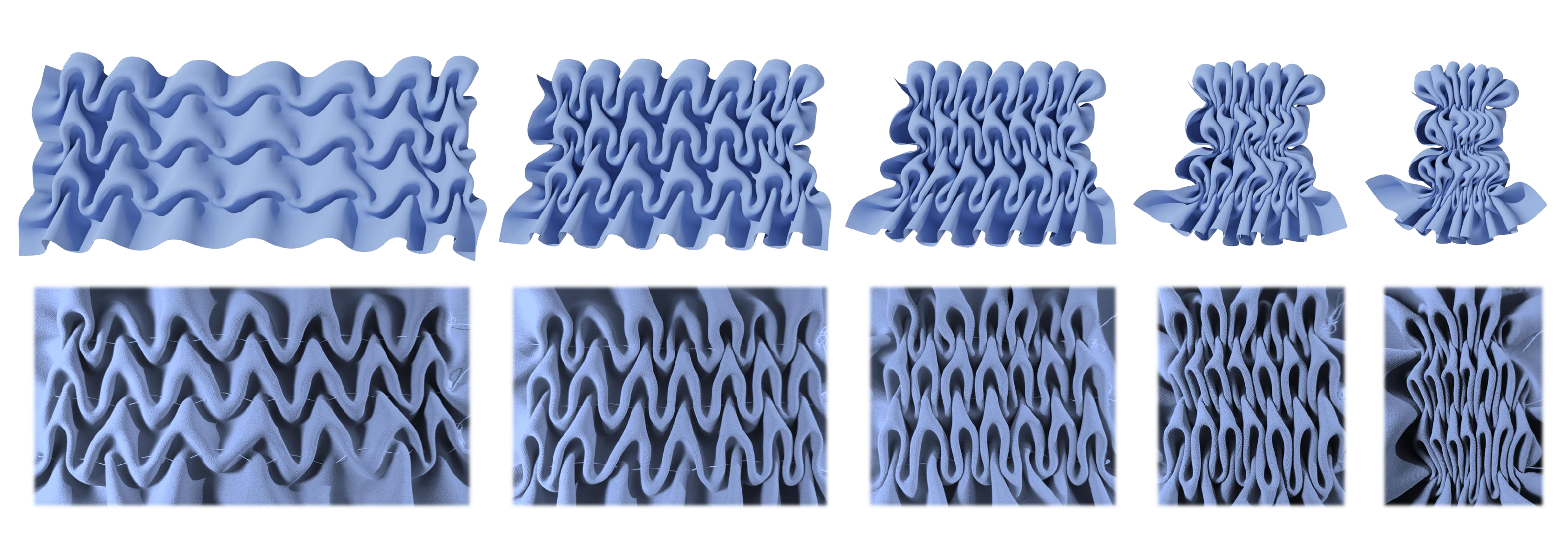}
    \put(13,30){\footnotesize $\gamma=50\%$}
    \put(40,30){\footnotesize $\gamma=30\%$}
    \put(61,30){\footnotesize $\gamma=20\%$}
    \put(77,30){\footnotesize $\gamma=10\%$}
    \put(91,30){\footnotesize $\gamma=5\%$}
    \end{overpic}
    \caption{Our results with different values of shrinkage $\gamma$ (\textbf{top}) and the corresponding fabrications (\textbf{bottom}).}
    \label{fig:res:zigzag-shrinkage}
\end{figure*}

\begin{figure*}[!t]
    \begin{overpic}[trim=-1.5cm 2.5cm 4.4cm 2cm,clip,width=1\linewidth,grid=false]{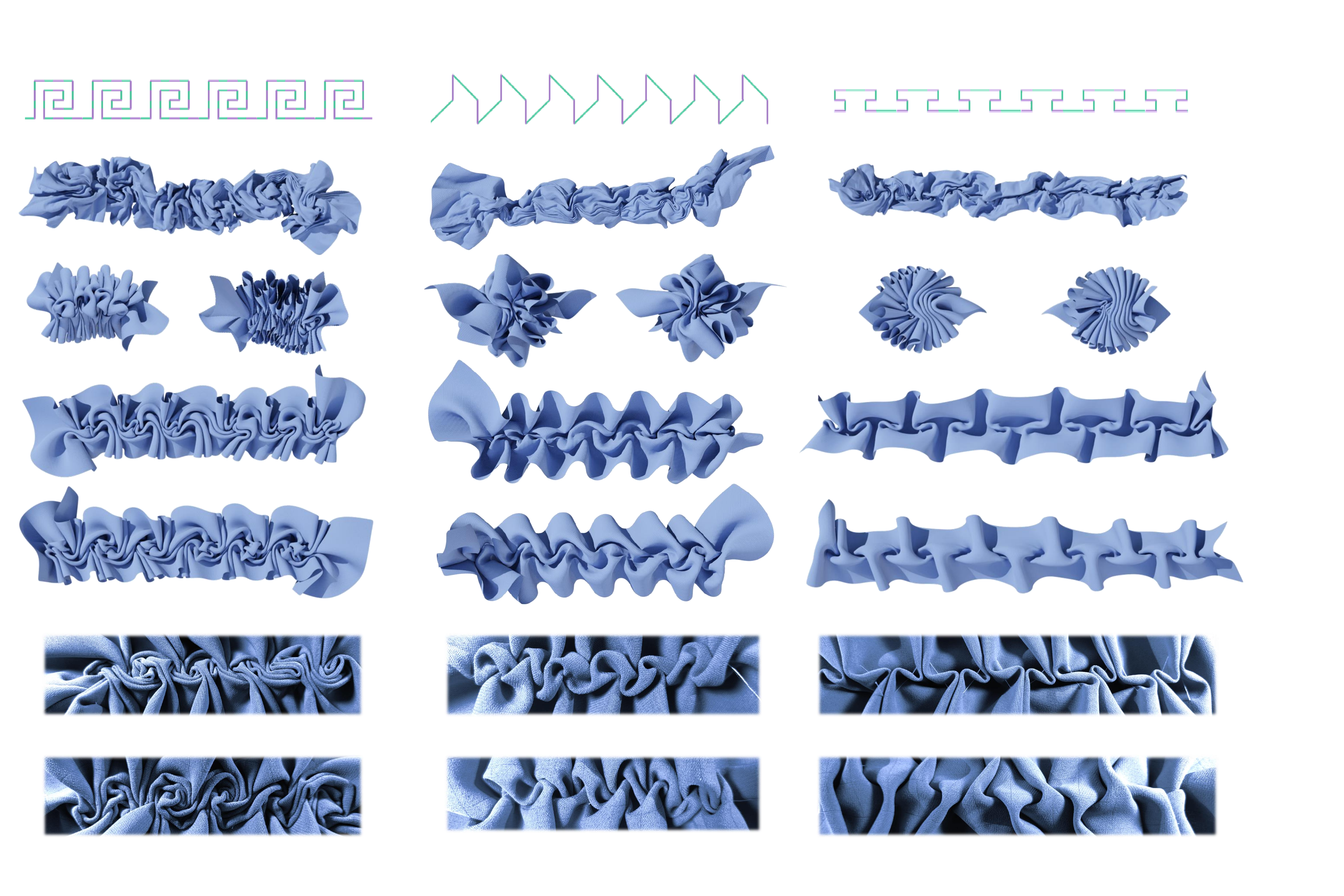}
    \put(-3,57){\footnotesize (a) patterns}
    % \put(-2,60){\footnotesize patterns}

    \put(-3,50){\footnotesize (b) Blender}
    \put(-1.5,48){\footnotesize (front)}

    \put(-3,41.5){\footnotesize (c) {\cipc}}
    \put(-2.5,39.5){\footnotesize \new{w/o priors}}
    \put(-4,37.5){\footnotesize (front \& back)}

    \put(-2,32.5){\footnotesize (d) \textbf{ours}}
    \put(-1,30.5){\footnotesize (front)}

    \put(-2,22.5){\footnotesize (e) \textbf{ours}}
    \put(-1,20.5){\footnotesize (back)}

    \put(-2,12.5){\footnotesize (f) \textbf{fabric}}
    \put(-1,10.5){\footnotesize (front)}

    \put(-2,4){\footnotesize (g) \textbf{fabric}}
    \put(-1,2){\footnotesize (back)}

    \put(14,26.5){\footnotesize $\gamma=10\%$}
    \put(46,26.5){\footnotesize $\gamma=20\%$}
    \put(80,26.5){\footnotesize $\gamma=30\%$}
    \end{overpic}\vspace{-3pt}
    \caption{We compare our method to Blender~\cite{blender}, {\cipc}~\cite{li2021CIPC} \new{(without any priors)}, and physical fabrications.}
    \label{fig:res:single-row-patterns}\vspace{-3pt}
\end{figure*}

\begin{figure}[!b]
    \centering
    \begin{overpic}[trim=0cm 0.5cm 1cm 0cm,clip,width=1\linewidth,grid=false]{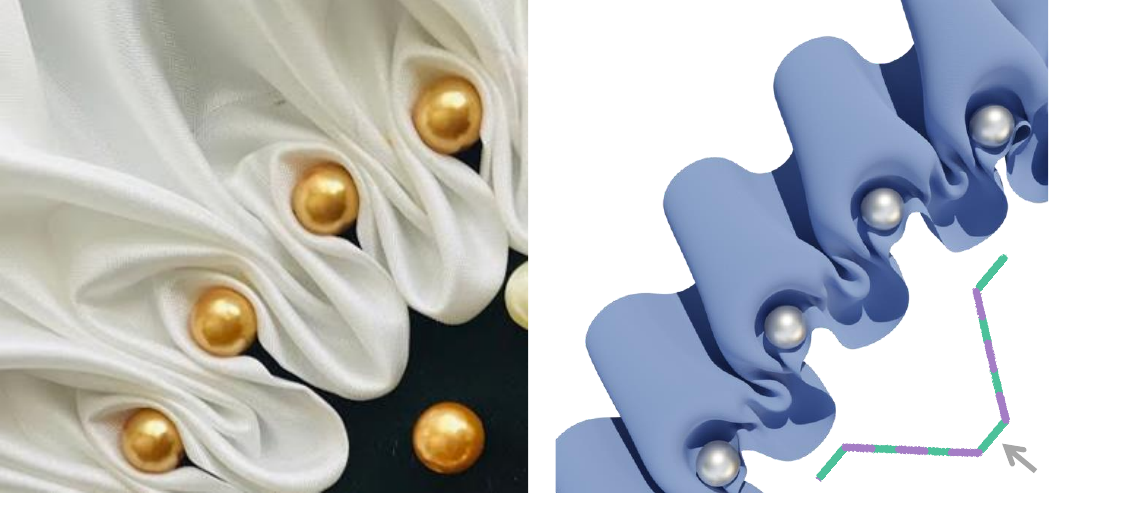}
\end{overpic}
\caption{Italian smocking with beads. \textbf{Left}: fabrication from \copyright FymsEmbroidery YouTube channel~\cite{egpearl}. Used with permission. \textbf{Right}:~our result.}
\label{fig:res:uneven-rest-length-ours}
\end{figure}

\section{Results}\label{sec:results}

\subsection{Comparisons to physical fabrications}
To validate that our method produces faithful results, we manually fabricated several Italian smocking patterns. 
Using a laser cutter, we engraved the pattern onto the fabric, followed by sewing through the fabric along the pattern rows and pulling the threads to achieve a similar appearance to that shown in YouTube tutorials. We then measured the approximate thread shrinkage $\gamma$ and executed our method using the same value.
In \figref{fig:res:multi-row-patterns} we show four examples of smocking patterns, where we color the front and back stitches in \textcolor{mygreen}{green} and \textcolor{mypurple}{purple}, respectively. 
We can see that our simulated results are close to the physical fabrications.

Italian smocking is relatively easy to fabricate, requiring merely the pulling of threads stitched through the fabric to create pleats. However, the complexity lies in pleat adjustment, as the resulting pleats after pulling are often irregular or distorted if the fabric does not properly slide along the threads (see \figref{fig:intro:fabrication-process}). The fabricator has to adjust the pleats to achieve the desired shape.
Therefore, in practice, designing a new Italian smocking pattern is challenging. If the fabricated result appears irregular or less pleasing, it is difficult to determine whether the issue lies in the pattern itself or the pleat refinement process was not adequately executed, since the pleat shape is unknown.
In \figref{fig:res:zigzag}, we showcase two patterns ((b) and (c)), not available online, and utilize our computed results to assist in adjusting the physically fabricated pleats to improve their shapes.
Specifically, we start from a known, classic pattern in \figref{fig:res:zigzag}~(a) and make local modifications to obtain the patterns (b) and (c), which retain the overall zigzag style. Despite the minor nature of these modifications, they lead to significant and unexpected changes in the final fabricated results. Our simulated results successfully helped us to adjust the shape of the physical pleats. We believe our algorithm can benefit artists in designing new patterns, as our results are consistently regular and reproducible in physical fabrications.

% \nf{In \figref{fig:res:zigzag}, starting with the classic Italian smocking pattern (\figref{fig:res:zigzag}(a)), we make local adjustments to some stitches, resulting in the patterns shown in \figref{fig:res:zigzag}(b) and \figref{fig:res:zigzag}(c). Despite the minor nature of these modifications, they lead to significant and unexpected changes in the final smocked results as observed in the real fabrications. However, our simulated results faithfully and precisely capture these variations and generate highly regular textures. This adaptability can benefit artists in designing new patterns and refining pleats to improve shapes without the need for time-consuming trial-and-error, as our results are consistently regular and reproducible in real fabrications. }

% We conduct another experiment in \figref{fig:res:zigzag}: we start with a zigzag pattern found online (\figref{fig:res:zigzag}(a)) and make local modifications to some of the front or back stitches, resulting in the patterns shown in \figref{fig:res:zigzag}(b-d).
% Despite the modifications, the overall stitching lines retain the zigzag style. However, as observed in the real fabrications, these local modifications lead to significant and unexpected changes in the final smocked results. 
% This demonstrates that our algorithm can benefit artists in pattern design and fine-tuning of spacing without the need for a time-consuming trial-and-error process, since our results consistently align with the real fabrications. 

\subsection{Justifications of formulation}
A key factor of our formulation is to model the gradual extraction of the thread during the fabric-thread interaction.
As detailed in \eqnref{eq:mtd:2d:prob:stitching-spring}, we achieve this by introducing the parameter $\gamma$ to describe the shrinkage of the total length of the threads. In \figref{fig:res:zigzag-shrinkage}, we compare our results when using different values of $\gamma$ to physical fabrications. 
It is evident that our results faithfully capture the smocking patterns, validating our choice of the parameter $\gamma$.

% \setlength{\columnsep}{10pt}%
% \setlength{\intextsep}{0pt}%
% \begin{wrapfigure}{r}{0.42\linewidth}
% \centering
% \begin{overpic}[trim=0cm 0cm 0cm 0cm,clip,width=1\linewidth,grid=false]{figures/eg_uneven_rest_length_ours.pdf}
% \end{overpic}
% \caption{Smock w/ beads (ours).}
% \label{fig:res:uneven-rest-length-ours}
% \end{wrapfigure}

% \begin{wrapfigure}[6]{r}{0.65\linewidth}
% \centering
% \vspace{-25pt}
% \begin{overpic}[trim=0cm 0.5cm 1cm 0cm,clip,width=1\linewidth,grid=false]{figures/eg_uneven_rest_length_ours_compare.pdf}
% \end{overpic}
% \caption{Smock w/ beads. \textbf{Left}: Real fabrication~\cite{egpearl}. \copyright Shagufta Fyms YouTube channel. Used with permission. \textbf{Right}: Our result.}
% \label{fig:res:uneven-rest-length-ours}
% \end{wrapfigure}
%Our formulation is flexible and capable of accommodating scenarios where the expected sewing distances are not evenly distributed. 
Our formulation is flexible and capable of accommodating scenarios where the expected sewing length distribution varies due to additional external constraints.
For example, artists may incorporate beads into the fabrication process (see \figref{fig:intro:fabrication-loose-thread}, left), which can be regarded as additional lower bound constraints for selected stitching springs. These constraints can be seamlessly integrated into our formulation, i.e., \eqnref{eq:mtd:2d:prob:stitching-spring}.
In \figref{fig:res:uneven-rest-length-ours}, we show an example of an Italian smocking pattern with beads. The arrow indicates the placement of the beads during fabrication.
For these selected stitching springs, we set the lower bound to the diameter of the beads. The resulting smocked design shows a distinct pleat pattern with an altered sewing length distribution imposed by these beads.

\subsection{Comparisons to cloth simulators}\label{sec:res:comparison}
\new{
\emph{Baselines.} We consider the closest setting in Blender~\cite{blender} (build-in cloth simulator), and {\cipc}~\cite{li2021CIPC} (vanilla version without any prior knowledge) as baselines, where two stitching points are expected to have zero distance after sewing and the fabric is not allowed to slide along the threads.
We admit that existing cloth simulators hold great potential to address the Italian smocking problem by modeling the threads with thin rods. However, this requires non-trivial considerations such as establishing boundary constraints and formulating pulling forces that guide the fabric to slide. 
We consider it a challenging open question and demonstrate our naive attempt at rod-fabric interaction in Appendix~\ref{app:rod-fab}.
}

\new{
\figref{fig:res:single-row-patterns} shows the comparison between our method, the considered baselines, and real fabrications.}
Specifically, we use the built-in cloth simulator in Blender to simulate the smocking process, treating each front and back stitch in the pattern as a sewing line. During the simulation, we manually halt the process when the total length of the sewing lines reaches $\gamma$ fraction of their initial lengths. 
We particularly fine-tune the maximum sewing force parameter to mimic the real fabrication process, ensuring that stitched threads are gently pulled out and avoiding rapid fabric shrinkage during simulation.
We can see that Blender fails to produce reasonable results: the solver cannot distinguish between the front and back stitches and fails to produce correct and consistent bending directions for the fabric during the simulation.
We also employ {\cipc} \new{without priors} to simulate smocking, treating each front and back stitch as a sewing line. However, the simulated results often appear cluttered due to {\cipc}'s assumption of zero-length sewing lines. Furthermore, without distinguishing between front and back stitching lines, the generated pleats exhibit randomly oriented bulges in both upward and downward directions.
In contrast, our preview results are more plausible and faithful in comparison with real fabrications.
% \new{TODO: add more discussion and experiments of rod-fabric setup.}
%\new{Besides, we also design an experiment of smocking with rod-fabric interaction. We consider it an open question and briefly discuss its complex setup in Appendix~\ref{app:rod-fab}.}

\begin{figure}[!t]
    \centering
    \begin{overpic}[trim=5.5cm 0.5cm 0cm -0.5cm,clip,width=0.98\linewidth,grid=false]
    %{figures/res4_canadian_run_italian.pdf}
    {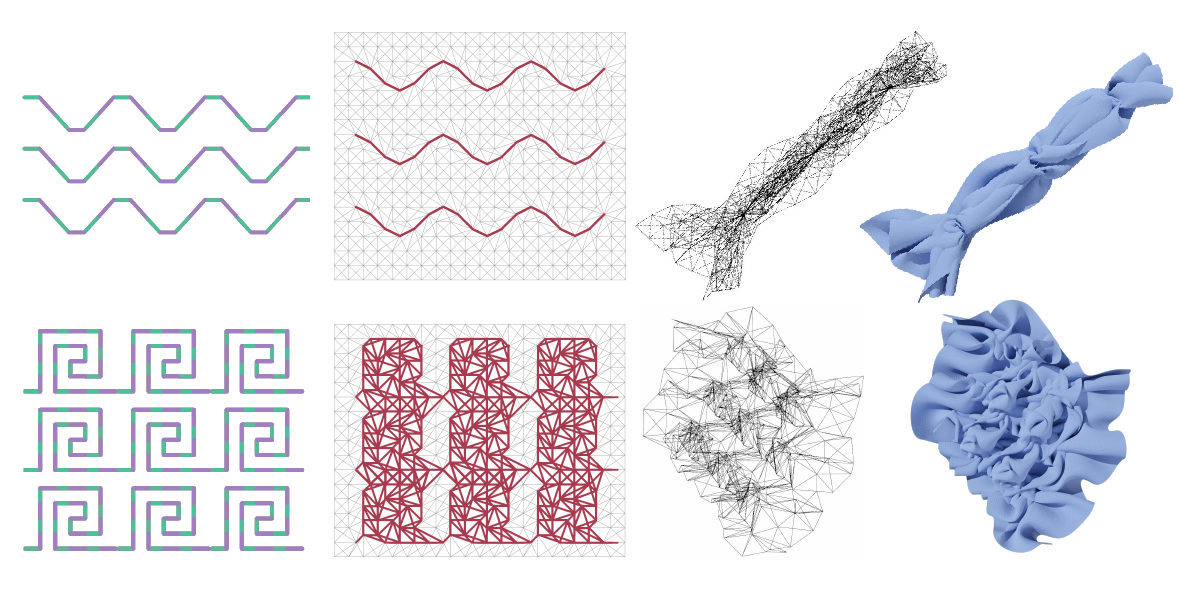}
    \put(7,65){\footnotesize smocked graph}
    \put(38,65){\footnotesize embedded graph (3D)}
    \put(78,65){\footnotesize \cite{ren2023smocking}}
    \end{overpic}
    \caption{\cite{ren2023smocking} fails to produce reasonable results on the Italian smocking patterns shown in \figref{fig:res:zigzag}~(a) and \figref{fig:res:single-row-patterns} (left).
    The underlay edges in the smocked graphs are highlighted in \textcolor{myred}{red}. }
    \label{fig:res:canadian-on-us}
\end{figure}

% \begin{figure}[!t]
%     \centering
%     \begin{overpic}[trim=0cm 0cm 0cm 0cm,clip,width=0.95\linewidth,grid=false]{figures/res5_ours_on_canadian.pdf}
%     \put(2,58){\footnotesize smocking pattern}
%     \put(41,58){\footnotesize \cite{ren2023smocking}}
%     \put(79,58){\footnotesize \textbf{ours}}
%     \end{overpic}\vspace{-3pt}
%     \caption{Our algorithm can be adapted to Canadian smocking. \new{TODO: physical fabrication result}}
%     \label{fig:res:ours-on-canadian}\vspace{-6pt}
% \end{figure}

\subsection{Adaptation for Canadian smocking}
\begin{figure}[!t]
    \centering
    \begin{overpic}[trim=0cm 0cm 0.5cm -1.2cm,clip,width=0.95\linewidth,grid=false]{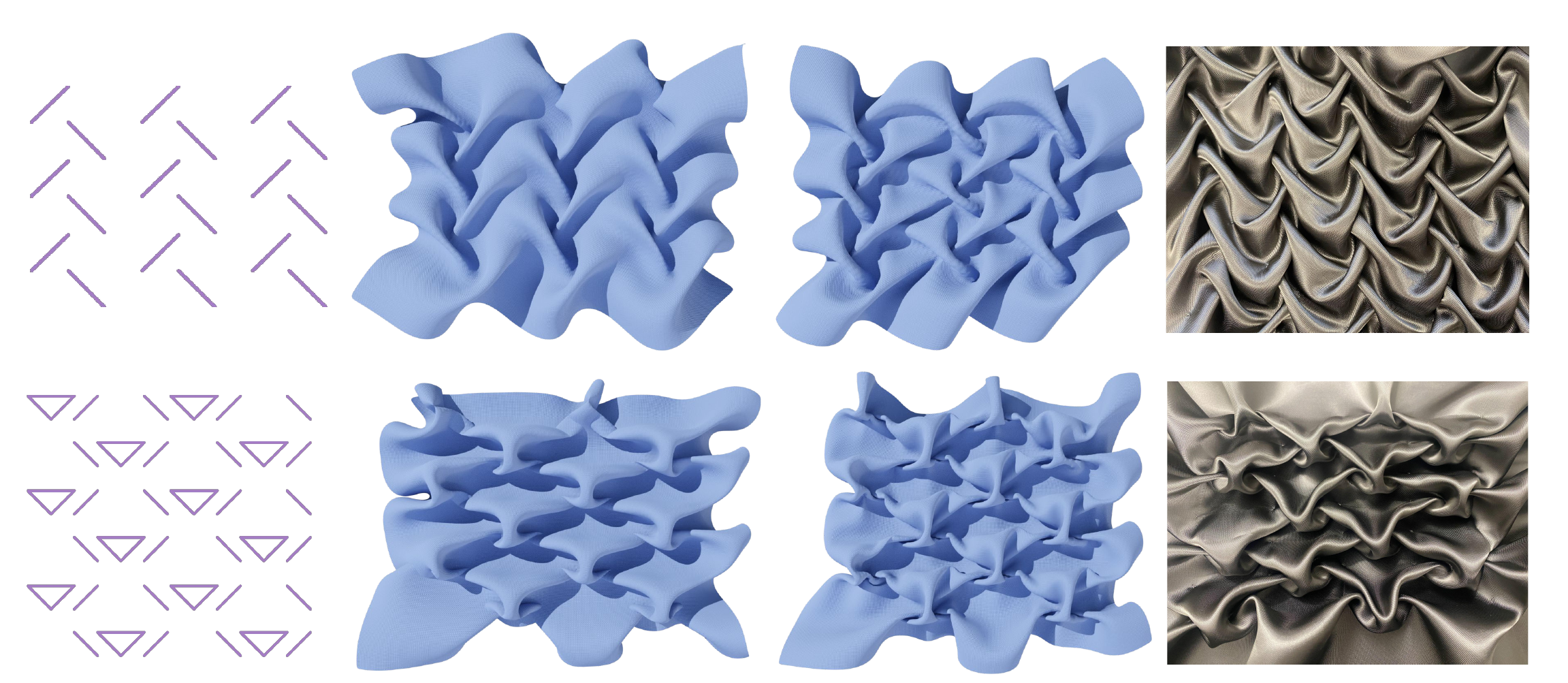}
    \put(5,43){\footnotesize pattern}
    \put(29,43){\footnotesize \cite{ren2023smocking}}
    \put(59,43){\footnotesize \textbf{ours}}
    \put(83,43){\footnotesize fabric}
    \end{overpic}\vspace{-3pt}
    \caption{\new{Our algorithm can be adapted to Canadian smocking. Physical fabrications are from~\cite{ren2023smocking}. Used with permission.}}
    \label{fig:res:ours-on-canadian}\vspace{-6pt}
\end{figure}

Applying the method designed for Canadian smocking~\cite{ren2023smocking} to Italian smocking leads to disastrous results, since all the stitching points on the stitching line are simply merged into a single node. We therefore decompose a stitching path $\ell=(v_1, v_2, \cdots, v_k)$ in an Italian smocking pattern into a set of separate stitching lines $\{\ell_1=(v_1,v_2), \ell_2 = (v_2, v_3),\cdots, \ell_{k-1}=(v_{k-1}, v_k)\}$ to avoid a degenerated smocked graph. However, even after this modification, the method in~\cite{ren2023smocking} still fails to produce reasonable results, as shown in \figref{fig:res:canadian-on-us}.
On the other hand, our method can be easily applied to Canadian smocking.
We define the set of back midpoints as empty ($\Vb = \emptyset$), set the expected lengths for all stitching springs to zero, and run our algorithm. See \figref{fig:res:ours-on-canadian} for some examples. Our method produces reasonable pleat shapes. 
Our results exhibit less regularity compared to~\cite{ren2023smocking}, primarily because their approach incorporates structural geometry priors, whereas our method is simulation-based and offers less control over the regularity of the pleats.

\begin{figure*}
    \centering
     \begin{overpic}[trim=0cm 0cm 0cm -1.8cm,clip,width=0.97\linewidth,grid=false]{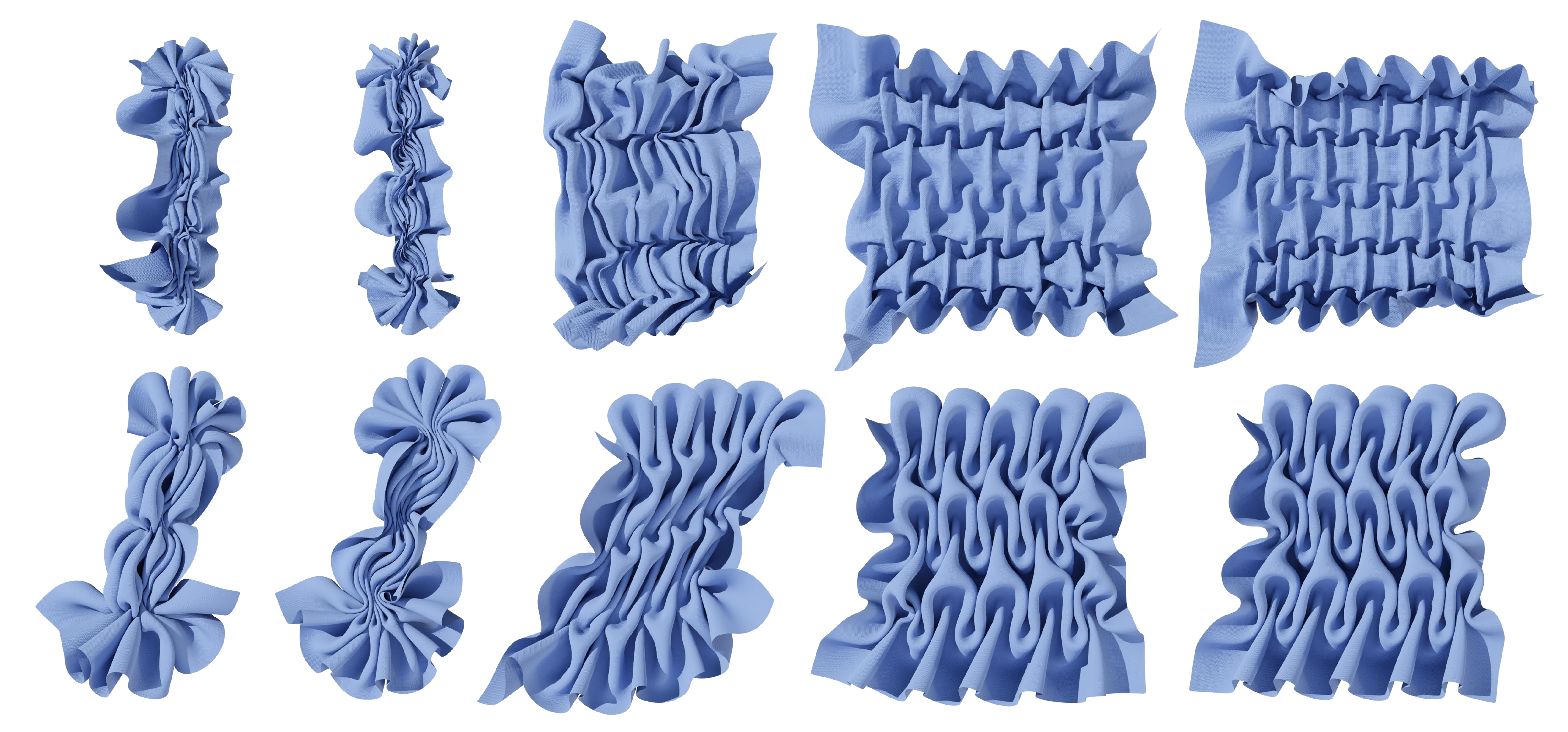}
     \put(8,47.5){\footnotesize  (a) \new{$E_{\text{cipc}}$}}
     \put(7.3,45.5){\footnotesize  \new{w/o priors}}
    
     \put(23,47.5){\footnotesize (b) \new{$E_{\text{cipc}}$}}
     \put(23,45.5){\footnotesize  \new{+ offset}}
    
     \put(36,47.5){\footnotesize (c) \new{$E_{\text{cipc}}$ + $w_sE_{\text{sew}}$}}
     \put(39,45.5){\footnotesize  \new{+ offset}}
     
     \put(57,47.5){\footnotesize (d) \new{$E_{\text{cipc}}$ + $w_pE_{\text{pos}}$}}
     \put(77,47.5){\footnotesize (e) \textbf{\new{$E_{\text{cipc}}$ + $w_sE_{\text{sew}}$ + $w_pE_{\text{pos}}$ }}}
     \put(85,44){\footnotesize $\gamma=30\%$}
     \put(85,23){\footnotesize $\gamma=20\%$}
     \end{overpic}\vspace{-6pt}
    \caption{\textbf{Ablation on geometric priors.} For two \new{patterns with the same unit pattern} shown in \figref{fig:res:multi-row-patterns}~(a) and \figref{fig:res:zigzag}~(a), we compare our results (shown in \textbf{(e)}) to four different settings: 
    \new{\textbf{(a)} {\cipc}~\cite{li2021CIPC} without any priors, \textbf{(b)} {\cipc} with offset initialization,} where we additionally provide initial configuration with inward/outward offset to guide {\cipc}, \textbf{(c)} we only provide the computed sewing lengths for {\cipc} i.e., $w_p = 0$, with offset initialization, \textbf{(d)} we only provide the computed positional constraints for {\cipc}, i.e., $w_s = 0$.
    }\label{fig:res:ablation}\vspace{-6pt}
\end{figure*}
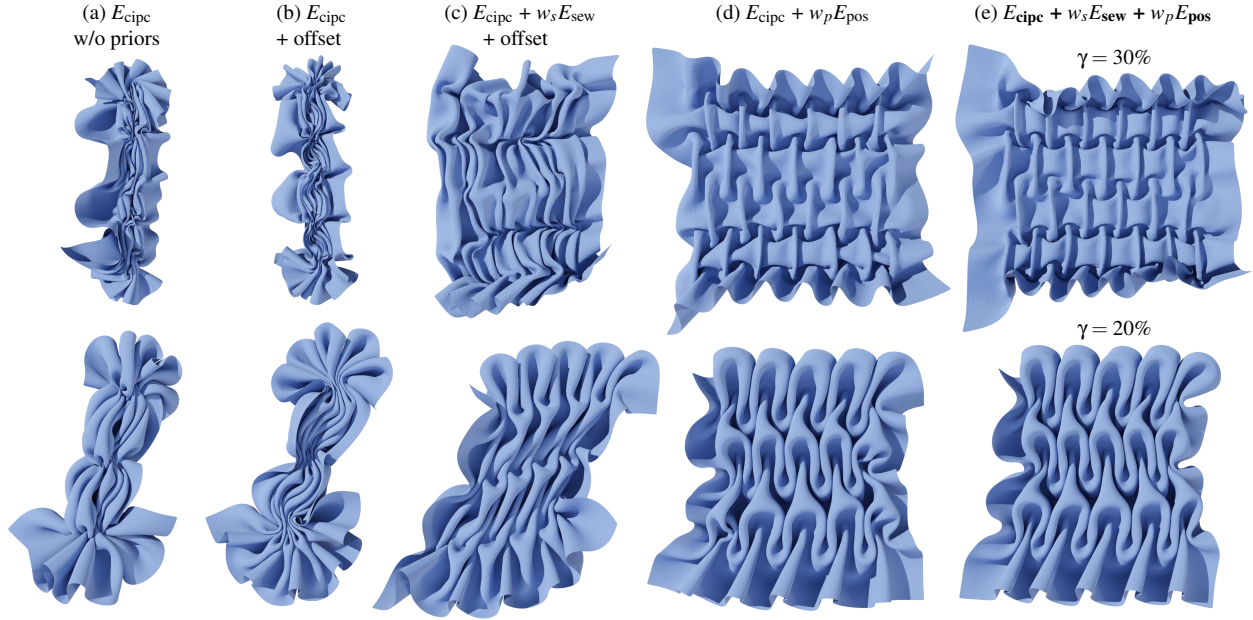

\subsection{Ablation on geometric priors}
In \figref{fig:res:ablation}, we conduct an ablation study to justify the usefulness of the computed sewing lengths and positions for guiding {\cipc}. In particular, we consider the following settings:
\begin{enumerate}[label={(\alph*)}]
    \item \new{$E_{\text{cipc}}$ without priors}: we use the released code of~\cite{li2021CIPC} for cloth draping with our fine-tuned parameters; the input stitching paths are considered standard sewing lines with an expected length of zero.
    \item \new{$E_{\text{cipc}}$ with offset initialization}: the midpoints in $\Vb$ (resp.\ $\Vf$) on the planar fabric mesh are offset with uniform positive (resp.\ negative) height. This modified mesh is then loaded in 
    {\cipc} as the first frame. This configuration guides the simulator to deform the mesh with the corresponding bending direction of front and back stitches.
    \item \new{$E_{\text{cipc}}+w_sE_{\text{sew}}$ with offset initialization}: setting $w_p=0$, we turn off the positional constraints and guide {\cipc} only use the computed sewing lengths. Here we use the initialization with offsets, since the sewing lengths do not contain the labeling of front and back stitches.
    \item \new{$E_{\text{cipc}}+w_pE_{\text{pos}}$}: similarly, we set $w_s = 0$ to analyze the importance of the positional constraints. 
    We do not use the offset initialization here, since our positional constraints already encode the front-back information.
    \item Ours, i.e., \new{$E_{\text{cipc}}+w_sE_{\text{sew}}+w_pE_{\text{pos}}$}. 
\end{enumerate}

When comparing (b) to (a), we can observe that the correct offsets in the initial configuration guide {\cipc} to create more regularly folded pleats. However, the overall result appears cluttered and excessively shrunken.
Incorporating the computed non-zero sewing lengths into {\cipc} results in less cluttered but still unrealistic pleats, since the sewing length constraints still permit significant deformation freedom. 
On the other hand, positional constraints offer a stronger prior for {\cipc} to achieve regular and realistic pleats.
Note that the stitching vertices and the sampled midpoints are treated equally in the positional constraints. Incorporating the sewing length regularizer can impose stricter constraints on the stitching vertices alone. This combined energy results in more regular pleats, particularly in the boundary region.

%\nf{While it can already generate regular patterns similar to our final results, we consider that the positional constraint treats the stitching vertices and the midpoints equally during convergence. 
%Therefore, we preserve the sewing distance energy to impose stricter constraints on the stitching vertices. 
%This combined approach, under the same convergence threshold, results in a slightly more realistic distribution of stitching force that helps interact with freely deformed regions. 
%As demonstrated in (d) and (e) in the top row, our results maintain flatter boundaries and pleats between different unit patterns. Also, the free boundaries of the pattern in the bottom row deform more naturally with these more accurate stitches.}

\subsection{Implementation, parameters \& runtime}\label{sec:res:implement}
\emph{Implementation.} We implement the 2D mass-spring system simulator in Python. The 3D deformer is implemented in C++ based on the released code of ~\cite{li2021CIPC} with modifications to incorporate non-zero sewing lengths and positional constraints. 
%The input smocking pattern is loaded as a triangle mesh in ``.obj'' format with front and back stitches stored as ``$l\quad v_1 \quad v_2$''.
%Our 2D simulator extracts the mass-spring system from the input and solves the dynamics, where the final sewing distances and the estimated 3D positions for  $v_i \in \Vs\cup\Vf\cup\Vb$ are exported and then loaded into our 3D deformer.
All the experiments are run on an Ubuntu system with an 8-core \unit[3.6]{GHz} Intel Core i7-9700K CPU and \unit[32]{GB} of RAM. 
\new{The full implementation can be found at \href{https://github.com/nifzhou/ItalianSmocking}{\webLink{https://github.com/nifzhou/ItalianSmocking}}.}
%We will release the data and code upon publication.

\noindent\emph{Parameters}.
The input smocking pattern is scaled to have the longest side of \unit[1]{m}.
For all experiments, we use the \emph{same} set of parameters: $w_s=0.1,\  w_p=0.01$.
The shrinkage parameter $\gamma$ is manually selected to replicate the textures demonstrated in online tutorials; we provide the values in the figures. 
%\new{We keep the simulation and material parameters unchanged to focus on the structural geometry of different smocking patterns.}
% \new{TODO: add explanations of why not adjusting the simulation/material parameter.}
Regarding the parameters for {\cipc}, we retain most of the parameters provided for cotton material in the released code of~\cite{li2021CIPC}. We set the bending stiffness scaling parameter to 100 and use the isotropic membrane model without strain limiting.
Our 3D deformer is executed for $100$ frames with a timestep size of $\Delta t=\unit[0.04]{sec}$, which is sufficient for convergence, starting from the fabric's initial planar state.
%Since the default {\cipc} continues to bend the mesh, running until convergence leads to cluttered results. We therefore manually select the intermediate frame that shows the best result without self-occlusion. For example, for \figref{fig:res:ablation}(b), we show the result at the 50th (top) and 60th frame (bottom), respectively.
%\jr{The pulling direction $\matr{d}$, discussed in \eqnref{eq:mtd:2d:pull-direction}, is a horizontal unit vector from left to right. 
%The fabric thickness $\tau$ is set to $0.01$ for all patterns except the \textsc{Curve} pattern in \figref{fig:mtd:curve-diff-shrinkage}. In this case, where the fabric crinkles sharply along highly rotational stitching paths, we adjust the thickness $\tau$ to $0.1$ to better capture locally squeezed multi-layer folds, leading to more realistic results.}
For all the patterns except the \textsc{Curve} pattern shown in \figref{fig:res:single-row-patterns} (left), we set the pulling direction (orthogonal to $\matr{d}$ discussed in \eqnref{eq:mtd:2d:pull-direction}), to a unit horizontal vector, and the fabric thickness $\tau$ to $0.01$. Altering the fabric thickness does not affect the appearance of the pleats, as these patterns rely on loose stitches to create their textures. 
The \textsc{Curve} pattern is more challenging, since the fabric crinkles sharply along the curly stitching paths, whose accumulative rotation is more than $2\pi$. 
% The local pulling direction varies for each front and back stitch. 
\new{The directions of stitching lines change significantly along the path and induce many local minima for the 2D spring system.}
We therefore align the pulling direction with each stitching spring instead of setting a uniform direction as in other patterns. We also adjust the thickness $\tau$ to $0.1$ to better capture locally squeezed multi-layer folds of the pattern size used in physical fabrication. 
%\nf{We also determine the pulling direction $\matr{d}$ discussed in \eqnref{eq:mtd:2d:pull-direction}) based on the cumulative degree of rotation along the stitching line within a unit pattern.
%direction $\matr{d}$ in Eq.~\eqref{eq:mtd:2d:pull-direction}.
% For simplicity, we can use the fabric thickness $\tau$ directly as the expected length for each stitching spring, thereby reducing computational overhead. \OSH{I don't get this last sentence, isn't the expected length the object of our optimization??} 
% \NF{For the \textsc{Curve} pattern with the complex rotation, $\matr{d}$ aligns with the direction of the stitching springs for each timestep, so instead of computing it each time, we can directly apply the real number (fabric thickness) here to accelerate the computation. (Though the fabric thickness is set to be higher than the others in order to generate results similar to real fabrication, which is $\tau=0.1$(for \textsc{Curve} pattern), $\tau=0.01$ (for other patterns) 

\begin{table}[!t]
    \caption{For the examples shown in \figreflist{fig:res:multi-row-patterns}{fig:res:zigzag}, we report the input complexity (including \#stitching vertices $\vert\Vs\vert$, \#grid vertices $\vert \V \vert$, \#vertices in the finer fabric mesh $n$, and shrinkage $\gamma$), the runtime for 2D simulation, 3D deformer, and total process.}\label{tab:res:runtime-main-res}\centering
\footnotesize
{\def\arraystretch{1}\tabcolsep=0.58em
\begin{tabular}{cccccccc}
\toprule[1pt]
\multirow{2}{*}{\begin{tabular}[c]{@{}c@{}}\itshape smocking\\ 
\itshape pattern\end{tabular}} & \multicolumn{4}{c}{\itshape complexity} & \multicolumn{3}{c}{\itshape runtime (minutes)} \\\cmidrule[0.8pt](l){2-5}\cmidrule[0.8pt](l){6-8}
 &   $\vert\Vs\vert$       & $\vert \V \vert$       & $n$  & $\gamma$      & 2D sim.   & 3D sim.   & total  \\ \midrule[1pt]
Fig.~\ref{fig:res:multi-row-patterns} (a) & 174 & 304 & 9116 & 30$\%$ & 0.04 & 2.35 & 2.39 \\
\rowcolor{myblue!20}
Fig.~\ref{fig:res:multi-row-patterns} (b) & 116 & 288 & 8858 & 20$\%$ &0.06 & 2.49 & 2.55 \\
Fig.~\ref{fig:res:multi-row-patterns} (c) & 438 & 880 & 33649 & 20$\%$ &0.45 & 11.67 & 12.12\\
\rowcolor{myblue!20}
Fig.~\ref{fig:res:multi-row-patterns} (d) & 222 & 638 & 21798 & 20$\%$ &0.18 & 7.21 & 7.39 \\
Fig.~\ref{fig:res:zigzag} (a) & 172 & 1175 & 42778 & 20$\%$ & 1.77 & 11.92 & 13.69 \\
\rowcolor{myblue!20}
Fig.~\ref{fig:res:zigzag} (b) & 144 & 1000 & 35916 & 30$\%$ & 0.70 & 8.63 & 9.33 \\
Fig.~\ref{fig:res:zigzag} (c) & 172 & 1363 & 50689 & 20$\%$ & 2.23 & 16.04 & 18.27 \\
\bottomrule[1pt]
\end{tabular}
}

% \centering
% \footnotesize
% {\def\arraystretch{1}\tabcolsep=0.82em
% \begin{tabular}{ccccccc}
% \toprule[1pt]
% \multirow{2}{*}{\begin{tabular}[c]{@{}c@{}}\itshape smocking\\ 
% \itshape pattern\end{tabular}} & \multicolumn{3}{c}{\itshape complexity} & \multicolumn{3}{c}{\itshape runtime (minutes)} \\\cmidrule[0.8pt](l){2-4}\cmidrule[0.8pt](l){5-7}
%  &   $\vert\Vs\vert$       & $\vert \V \vert$       & $n$        & 2D sim.   & 3D sim.   & total  \\ \midrule[1pt]
% Fig.~\ref{fig:res:multi-row-patterns} (a) & 174 & 304 & 9116 & 0.04 & 2.35 & 2.39 \\
% \rowcolor{myblue!20}
% Fig.~\ref{fig:res:multi-row-patterns} (b) & 116 & 288 & 8858 & 0.06 & 2.49 & 2.55 \\
% Fig.~\ref{fig:res:multi-row-patterns} (c) & 438 & 880 & 33649 & 0.45 & 0012.4 & 12.9 \\
% \rowcolor{myblue!20}
% Fig.~\ref{fig:res:multi-row-patterns} (d) & 222 & 638 & 21798 & 0.18 & 7.21 & 7.39 \\
% Fig.~\ref{fig:res:zigzag} (a) & 172 & 1175 & 42778 & 1.77 & 11.92 & 13.69 \\
% \rowcolor{myblue!20}
% Fig.~\ref{fig:res:zigzag} (b) & 144 & 1000 & 35916 & 0.70 & 8.63 & 9.33 \\
% Fig.~\ref{fig:res:zigzag} (c) & 172 & 1363 & 50689 & 2.23 & 16.04 & 18.27 \\
% \bottomrule[1pt]
% \end{tabular}
% }

\end{table}

\begin{table}[!t]
    \caption{The total runtime for the ablation study in Fig.~\ref{fig:res:ablation}.}
    \label{tab:res:runtime-ablation}
    \centering
\footnotesize
{\def\arraystretch{1.1}\tabcolsep=0.34em
\begin{tabular}{cccccccccc}
\toprule[1pt]
\multirow{2}{*}{\begin{tabular}[c]{@{}c@{}}\itshape smocking\\ 
\itshape pattern\end{tabular}}  & \multicolumn{4}{c}{ \itshape complexity} & \multicolumn{5}{c}{ \itshape total runtime (minutes)} \\ \cmidrule[0.8pt](l){2-5}\cmidrule[0.8pt](l){6-10}
&   $\vert\Vs\vert$       & $\vert \V \vert$       & $n$    & $\gamma$    & (a)   & (b)   & (c)  & (d)  & (e) \textbf{ours}  \\ \midrule[1pt]
Fig.~\ref{fig:res:ablation} (a) & 135 & 252 & 7300 & 30$\%$ & 5.46  & 4.98  & 1.88  & 1.19  & 1.79  \\
% \rowcolor{myblue!20}
Fig.~\ref{fig:res:ablation} (b) & 93 & 700 & 24069 & 20$\%$ &16.75  & 14.33  & 8.83  & 6.19  & 6.52  \\ \bottomrule[1pt]
\end{tabular}
}

% \centering
% \footnotesize
% {\def\arraystretch{1.1}\tabcolsep=0.56em
% \begin{tabular}{ccccccccc}
% \toprule[1pt]
% \multirow{2}{*}{\begin{tabular}[c]{@{}c@{}}\itshape smocking\\ 
% \itshape pattern\end{tabular}}  & \multicolumn{3}{c}{ \itshape complexity} & \multicolumn{5}{c}{ \itshape total runtime (minutes)} \\ \cmidrule[0.8pt](l){2-4}\cmidrule[0.8pt](l){5-9}
% &   $\vert\Vs\vert$       & $\vert \V \vert$       & $n$        & (a)   & (b)   & (c)  & (d)  & (e) \textbf{ours}  \\ \midrule[1pt]
% Fig.~\ref{fig:res:ablation} (a) & 135 & 252 & 7300 & 5.46  & 4.98  & 1.88  & 1.19  & 1.79  \\
% \rowcolor{myblue!20}
% Fig.~\ref{fig:res:ablation} (b) & 93 & 700 & 24069 & 16.75  & 14.33  & 8.83  & 6.19  & 6.52  \\ \bottomrule[1pt]
% \end{tabular}
% }
\end{table}

\noindent\emph{Runtime.}  
In Table~\ref{tab:res:runtime-main-res}, we report the smocking pattern complexity and the runtime breakdown of our algorithm for the examples shown in~\figreflist{fig:res:multi-row-patterns}{fig:res:zigzag}. 
%This includes the number of stitching vertices ($\vert\Vs\vert$), grid vertices ($\vert\V\vert$), the vertices ($n$) in the finer fabric mesh, and the shrinkage ($\gamma$).
The runtime of the 2D simulator depends on the number of grid vertices ($\vert\V\vert$), the number of stitching vertices ($\vert\Vs\vert$), and the shrinkage ($\gamma$).
The runtime of the 3D deformer depends on the number of vertices in the finer fabric mesh ($n$) and the number of constraints added to {\cipc}, closely related to $\vert\Vs\vert$.

\setlength{\columnsep}{3pt}%
\setlength{\intextsep}{0pt}%
\begin{wrapfigure}{r}{0.48\linewidth}
\centering
\begin{overpic}[trim=0cm 0cm 0cm 0cm,clip,width=1\linewidth,grid=false]{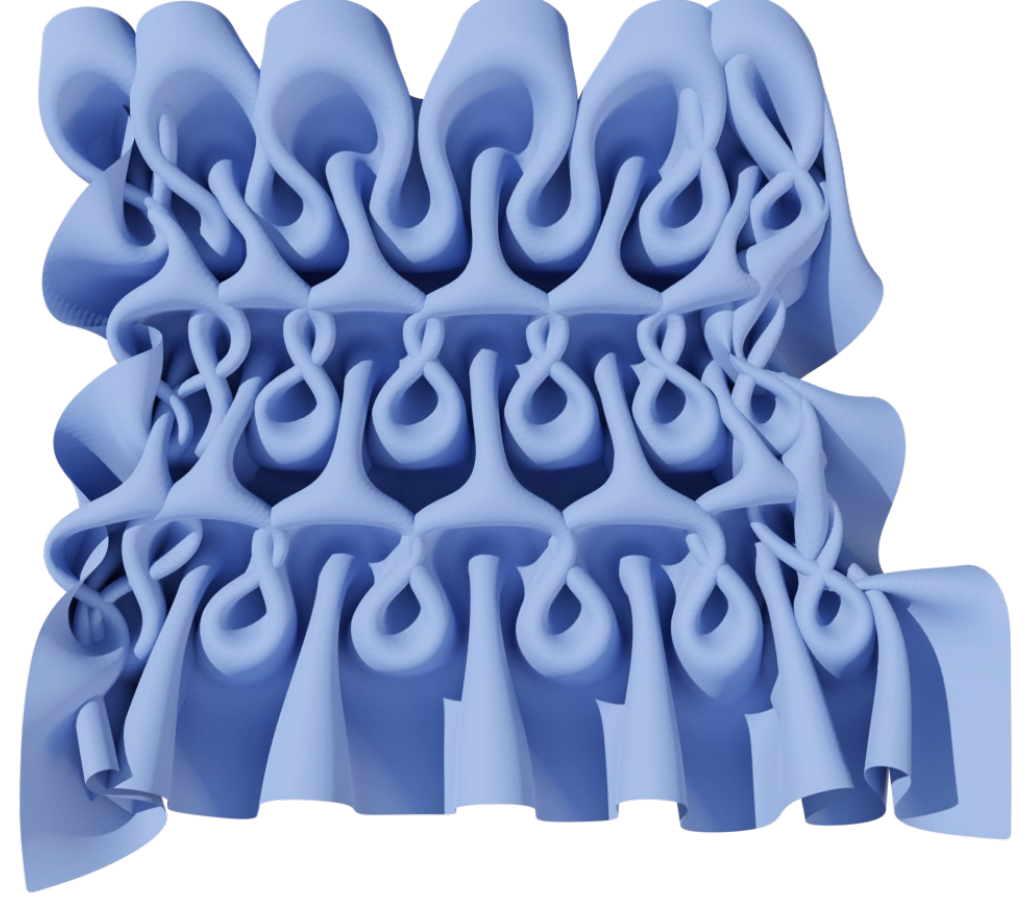}
\put(10,-2){\footnotesize $t_\text{arap} = \unit[13]{sec},\ t_\text{cipc} = \unit[536]{sec}$}
\end{overpic}
\caption{Preview via {\arap}.}
\label{fig:res:zigzag-arap}
\end{wrapfigure}
Our 2D simulator is highly efficient, taking only seconds to converge even for smocking patterns with hundreds of stitching vertices. The 3D deformer can take minutes to deform the finer representation of the fabric with 10-50K vertices with
collision handling. 
We also implemented a fast preview algorithm based on {\arap}~\cite{arap} with our computed positional constraints incorporated. 
% \new{TODO: mention using ARAP without material consideration.}
This method is much more efficient and only takes a few seconds to preview the smocked pleats \new{without material consideration}. However, since there is no collision handling in {\arap}, when the shrinkage is significant, the final result can have obvious self-intersections, as shown in \figref{fig:res:zigzag-arap}
\new{ (see our result in \figref{fig:res:zigzag-shrinkage} ($\gamma=20\%$))}. 

The {\cipc}-based deformer can handle self-collisions and produce more visually pleasing results, albeit at a higher computational cost. 
In practice, users or designers can use the {\arap}-based deformer for a quick preview of the smocking results and switch to the {\cipc}-based deformer to generate final results once satisfied with the smocking pattern design.

It is worth noting that incorporating additional non-zero sewing lengths and positional constraints actually speeds up the 3D simulation, contrary to intuition. In Table~\ref{tab:res:runtime-ablation} we report the runtime comparison of the ablation study in Fig.~\ref{fig:res:ablation}, where we run each setting for 100 frames for a fair comparison.
Our method, setting (e), is more than twice as fast as the original {\cipc} without constraints (setting (a)). 
We believe this is because our accurate constraints reduce the search space, leading to faster convergence.
Similar behavior is observed when comparing the setting (b) to (a), where adding the correct offset to the initialization helps {\cipc} converge quicker.

\section{Conclusion, limitations, \& future work}\label{sec:conclusion}
In this work, we formalize the Italian smocking embroidery technique and propose a simple method to simulate the smocking results. 
Unlike Canadian smocking, Italian smocking involves continuous stitching lines that traverse the fabric. 
Pulling the free ends of the threads to gather the fabric along the stitching paths gives rise to intricate and complex pleat patterns.
The distances between two stitching vertices in the final result are not necessarily zero, which makes existing methods, \new{with zero-length stitching primitives,} not directly suitable for the task.
Our method consists of two main steps: firstly, we model the fabric as a coarse mass-spring system and solve for its projected 2D configuration by iteratively estimating the expected embedding length of fabric springs and stitching springs. Next, we incorporate the computed sewing lengths, along with the positions of the stitching vertices and their midpoints, into the {\cipc} simulator to guide fabric deformation in finer resolution.
Our method achieves more faithful results compared to baseline approaches. 
It is important to highlight that our smocked results closely resemble physical fabrications in structural and qualitative aspects, while not being completely identical. The inherent randomness in pleating persists, since the shape is not fully constrained, which is also a characteristic of smocking.

Our 2D simulator is efficient and offers accurate priors for 3D mesh deformation. However, running {\cipc} to address self-collisions while integrating the computed constraints is time-consuming due to the fine level of details in the resulting pleats.
A more efficient self-collision handling would greatly benefit interactive design, which we leave as future work.
Another limitation of our formulation is that we use the shrinkage parameter $\gamma$ with respect to the total length of threads to model the pulling process, implicitly assuming simultaneous shrinking for all stitches. 
% \new{TODO: explain a bit why simultaneous and why it may also lead to not identical unit pattern as in Fig.11.} 
In actual fabrication, it is common for stitches closer to the pulling end to shrink first, with the effect propagating to stitches farther away. While our final smocked results are reasonably faithful, the simulation process deviates from physical fabrication. It would be interesting to explore a more advanced and sophisticated formulation for accurately simulating the pulling process.
%
% \new{TODO: discussion of potential extension to the curved surface for computer-aided garment design/textile design/virtual prototyping.}
\new{In addition, we currently only address smocking in a planar configuration. 
Future work involving extensions to curved surfaces, such as adapting positional constraints through parameterization lifting, would  contribute to digital textile design, providing realistic smocking patterns.}
Another interesting direction for future work is to develop an interactive system for pleated fabric. 
Many online smocking tutorials necessitate pleat adjustment by the fabricator to achieve the desired results, as pulling threads can result in irregular pleats. A digital equivalent in the form of an interactive system for digital design would be valuable.

\section*{Acknowledgments}
We would like to thank the anonymous reviewers for their insightful feedback.
We extend our gratitude to 
M.\ Rifad (YouTube channel ``DIY Stitching''), 
F.\ Shanas (YouTube channel ``handiworks''), 
and S.\ Fyms (YouTube channel ``FymsEmbroidery'') 
for generously granting us permission to use the images of their remarkable fabrication results.
This work was supported in part by the ERC Consolidator Grant No.\ 101003104 (MYCLOTH).

%-------------------------------------------------------------------------
% bibtex
\bibliographystyle{eg-alpha-doi} 
\bibliography{egbibsample}       

% biblatex with biber
% \printbibliography                

%-------------------------------------------------------------------------

\appendix
\begin{figure}[!b]
    \centering
    \begin{overpic}[trim=0cm 0.9cm 0cm 0cm,clip,width=1\linewidth,grid=false]{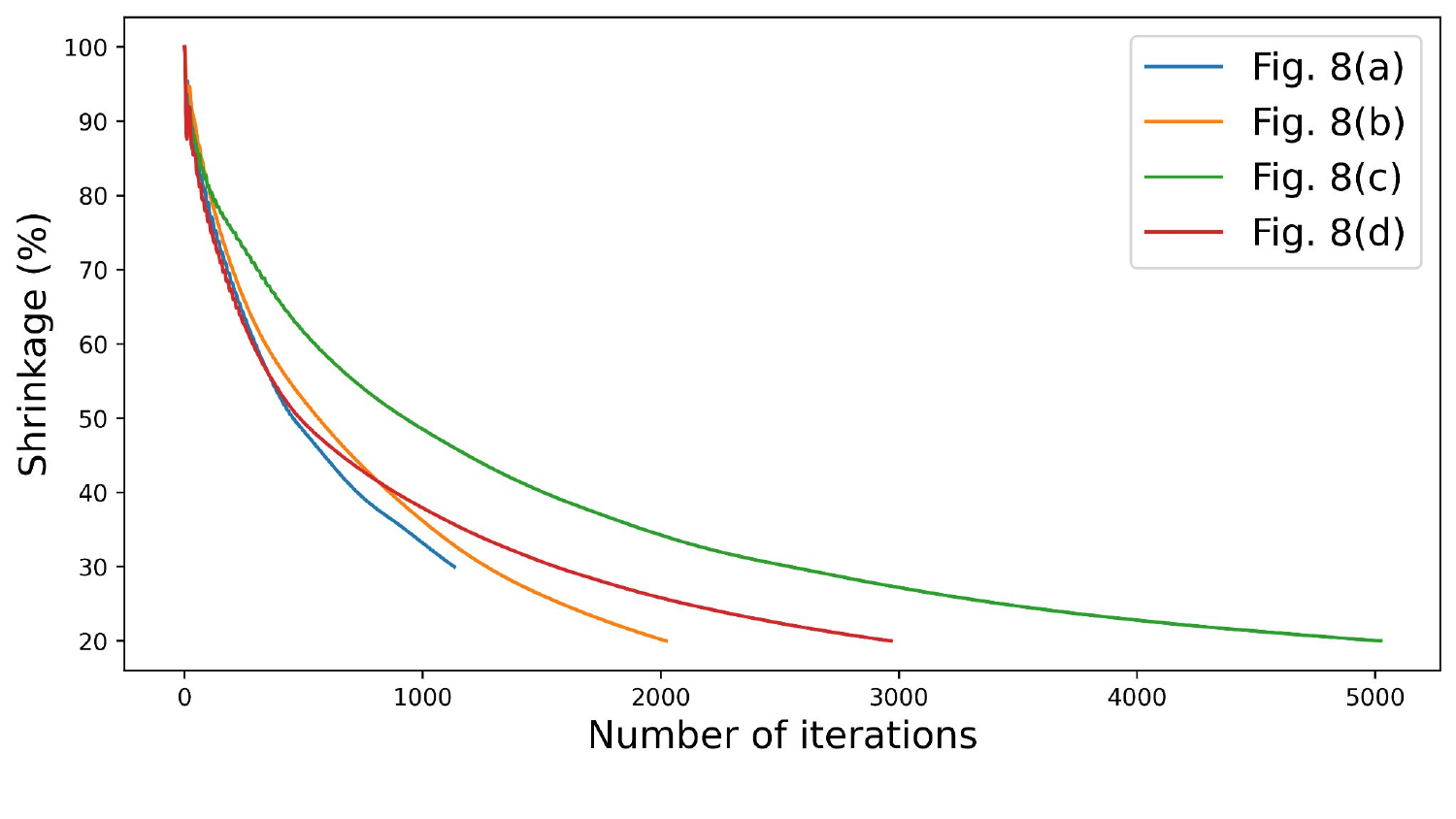}
    \end{overpic}\vspace{-12pt}
    \caption{\new{Convergence analysis of Alg.~\ref{algo:2d-sim} for the patterns in \figref{fig:res:multi-row-patterns}.}}
    \label{fig:appx:convergence}
\end{figure}

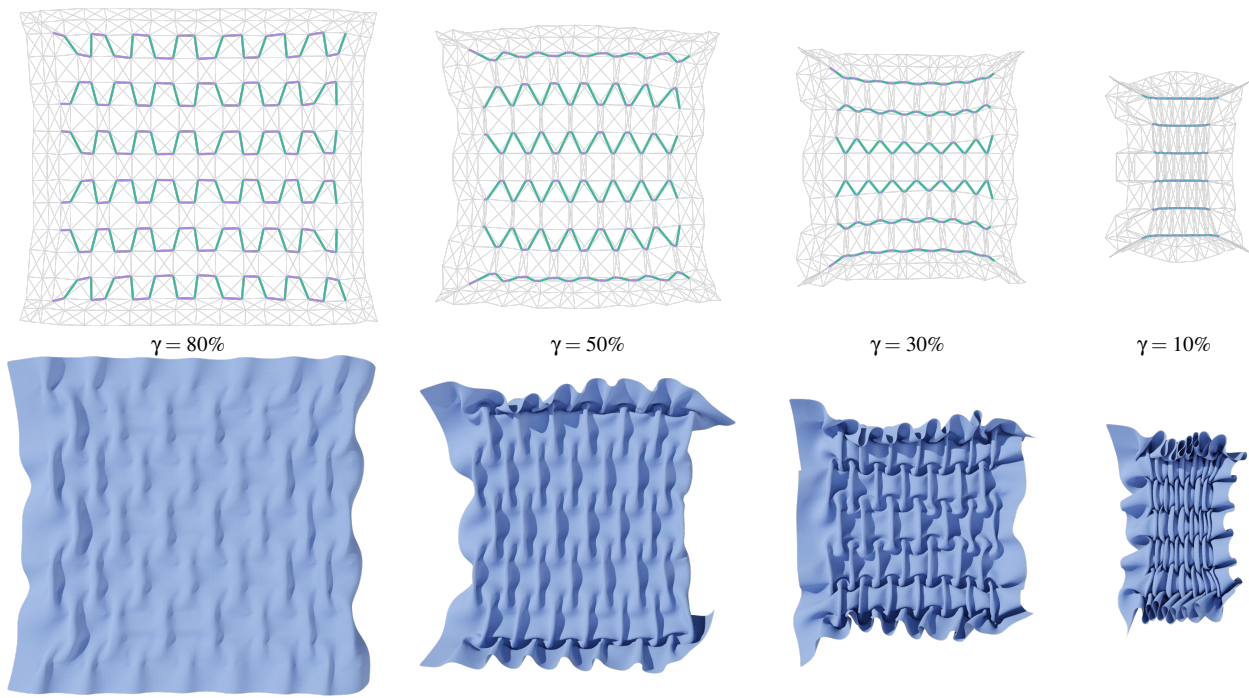
\begin{figure*}[!t]
    \centering
    \begin{overpic}[trim=0cm 0cm 0cm -0.5cm,clip,width=1\linewidth,grid=false]{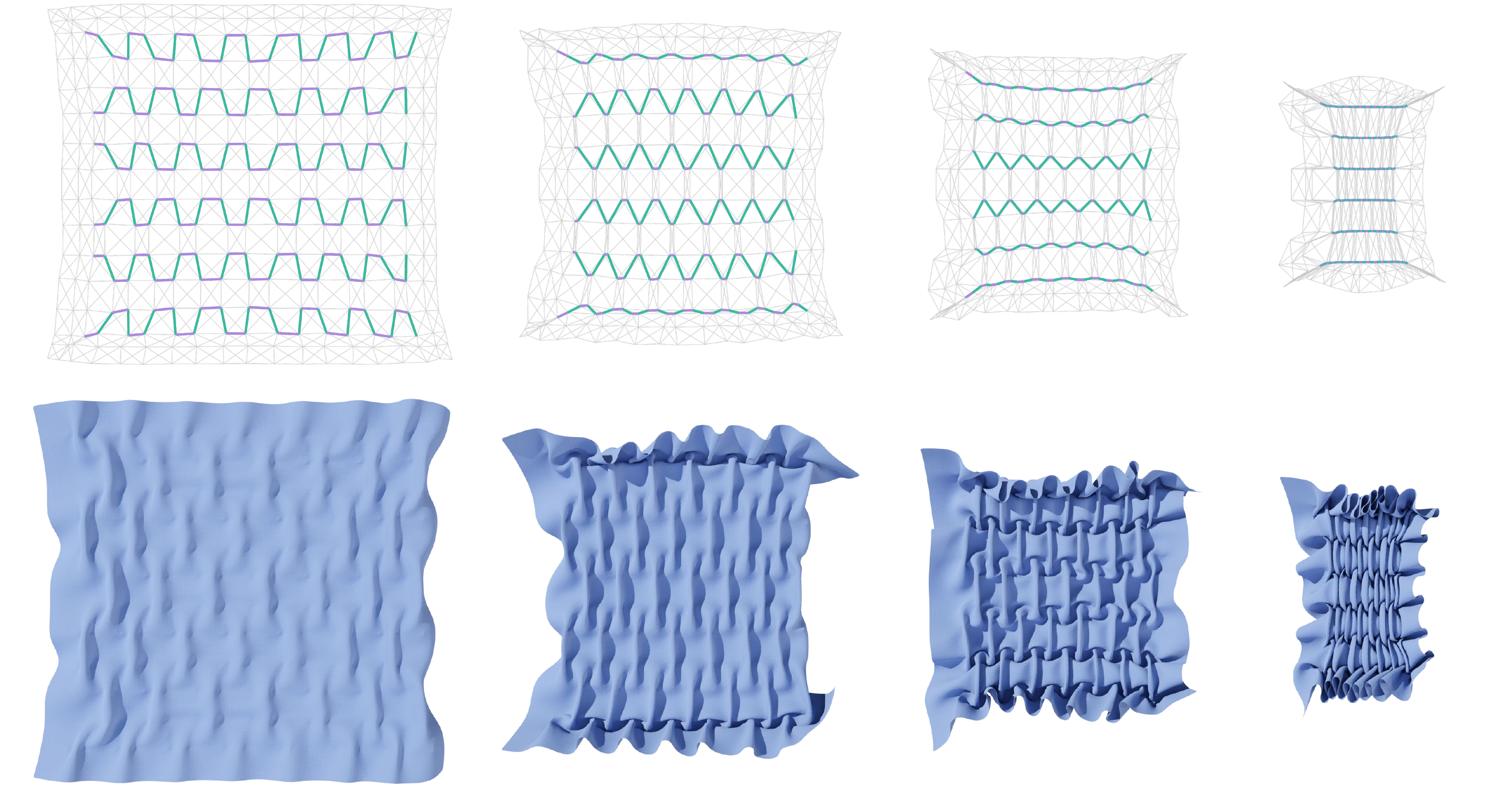}
    % \put(-1,40){\footnotesize (a) 2D}
    % \put(-2.5,38){\footnotesize embedding}
    % \put(-1,12){\footnotesize (b) 3D}
    % \put(-0.5,10){\footnotesize mesh}
    \put(13,26){\footnotesize $\gamma=80\%$}
    \put(43,26){\footnotesize $\gamma=50\%$}
    \put(67,26){\footnotesize $\gamma=30\%$}
    \put(87,26){\footnotesize $\gamma=10\%$}
    \end{overpic}\vspace{-3pt}
    \caption{\new{The 2D simulation (\emph{top}) and the corresponding 3D results (\emph{bottom}) with different shrinkages $\gamma$ for the pattern shown in \figref{fig:res:multi-row-patterns}~(a).}}
    \label{fig:appx:2d_iter}
\end{figure*}

\section{\new{Algorithm convergence behavior}}\label{app:convergence}
Here we analyze the convergence of Algorithm~\ref{algo:2d-sim}, designed to address the constrained problem discussed in \secref{sec:mtd:2d}.
Specifically, the goal is to simulate the 2D mass-spring system with dynamic target spring lengths, adhering to the constraints detailed in \eqnref{eq:mtd:2d:prob:fabric-spring} and \eqnref{eq:mtd:2d:prob:stitching-spring}.
Algorithm~\ref{algo:2d-sim}, by design, successfully satisfies all the constraints except for the shrinkage requirement over iterations. 
Therefore, we consider the algorithm converged when the current shrinkage ratio aligns with the user-specified target shrinkage.

% \emph{Problem.} The problem is formed by a 2D mass-spring system with dynamic expected spring lengths. The conditions, shown in \eqnref{eq:mtd:2d:prob:fabric-spring} and \eqnref{eq:mtd:2d:prob:stitching-spring}, should be met to reach its convergence. Constraints except the shrinkage requirement are initially met and kept inviolated throughout the simulation. Thus, we consider the current shrinkage as the convergence measure.
% problem, stopping criterion, convergence measure

In \figref{fig:appx:convergence} we report the convergence results for the four patterns showed in \figref{fig:res:multi-row-patterns}.
The convergence behavior exhibits a similar logarithmic decrease across different patterns.
Notably, the green line in \figref{fig:appx:convergence}, corresponding to the pattern in \figref{fig:res:multi-row-patterns}~(c),
shows a slower convergence. 
This is attributed to the vertical stitching lines, which tend to shrink less rapidly compared to stitching lines in other orientations when the threads are pulled horizontally. This behavior aligns with real fabrications and is explicitly considered in \eqnref{eq:mtd:2d:pull-direction}.

% \emph{Result.} The convergence behavior shows a similar logarithmic decreasing trend across patterns. See \figref{fig:appx:convergence} for the convergence plot of the example patterns illustrated in \figref{fig:res:multi-row-patterns}. The pattern in \figref{fig:res:multi-row-patterns}~(c) with many vertical stitching lines shows a slower convergence, which results from the assumption of larger expected spring lengths in \eqnref{eq:mtd:2d:pull-direction}. 
% For intuitive visualization of the 2D simulated system, we illustrate several results of an example pattern under different shrinkages in \figref{fig:appx:2d_iter}.
% Since shrinkage is the only violated constraint during simulation, it generates valid positional constraints with current shrinkage at each step.

\figref{fig:appx:2d_iter} shows a qualitative example showcasing the algorithm's progression at different shrinkage values and the corresponding 3D results derived from the solved 2D configurations.
The intermediate steps can still provide valid positional constraints to achieve realistic smocked results in 3D before the target shrinkage is reached.  
This feature benefits simulation of the shrinking process under different shrinkages in a single pass, as shown in \figref{fig:res:zigzag-shrinkage} and \figref{fig:appx:2d_iter}.

% \JR{add convergence behavior here}
% \JR{add one 2D mass-spring over iteration illustration}

\begin{figure*}[!t]
    \centering
    \begin{overpic}[trim=0cm 0cm 0cm -0.5cm,clip,width=0.95\linewidth,grid=false]{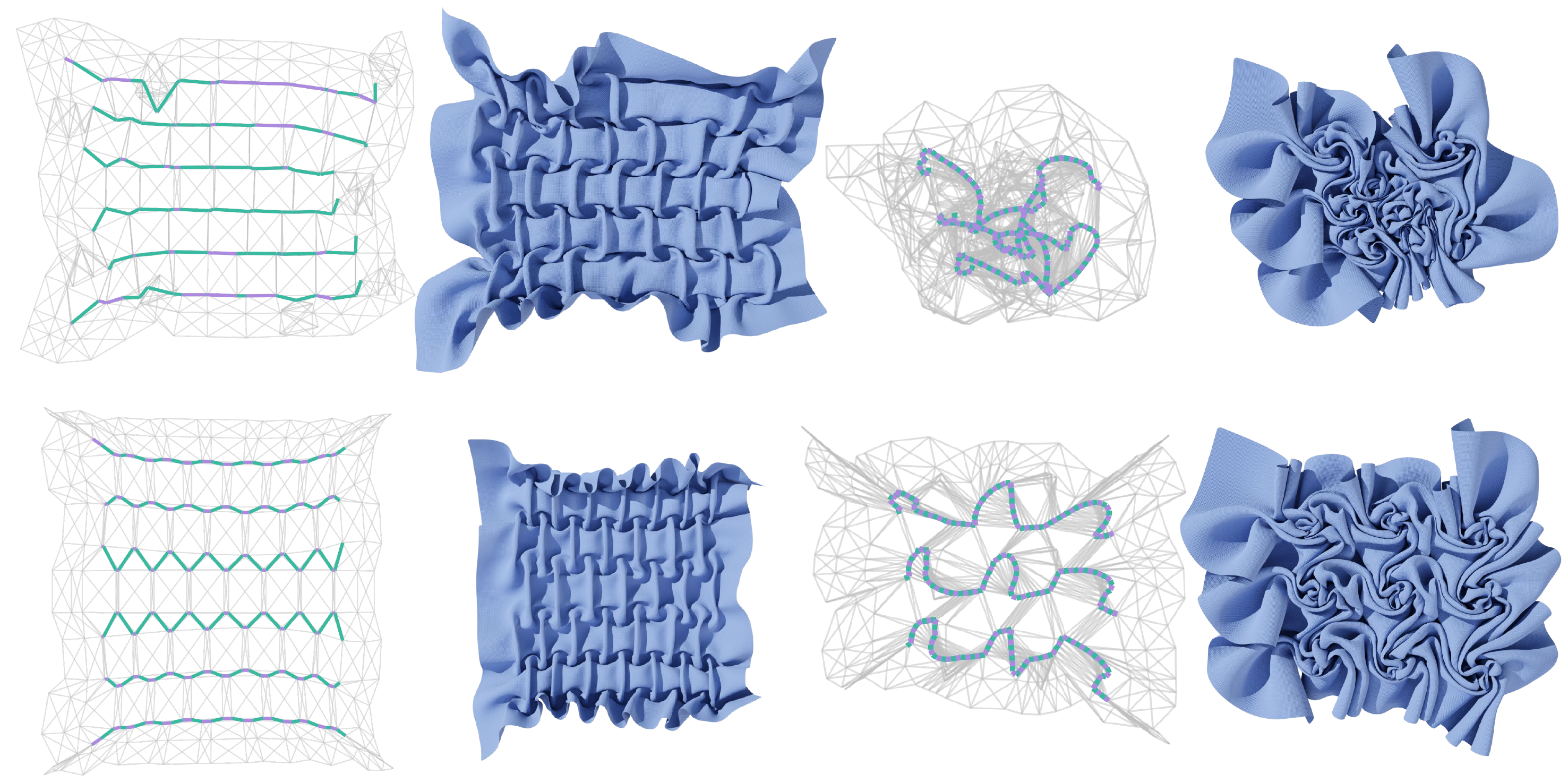}
    \put(-5,38){\footnotesize (a) SLSQP}
    \put(-4,10){\footnotesize (b) \textbf{ours}}
    \put(24,24.5){\footnotesize $\gamma=30\%$}
    \put(73,24.5){\footnotesize $\gamma=10\%$}
    \end{overpic}\vspace{-3pt}
    \caption{\new{We compare our simulated results to SQP-optimized results on two patterns shown in \figref{fig:res:multi-row-patterns}~(a) (left) and \figref{fig:teaser} (right).}}
    \label{fig:appx:sqp}
\end{figure*}
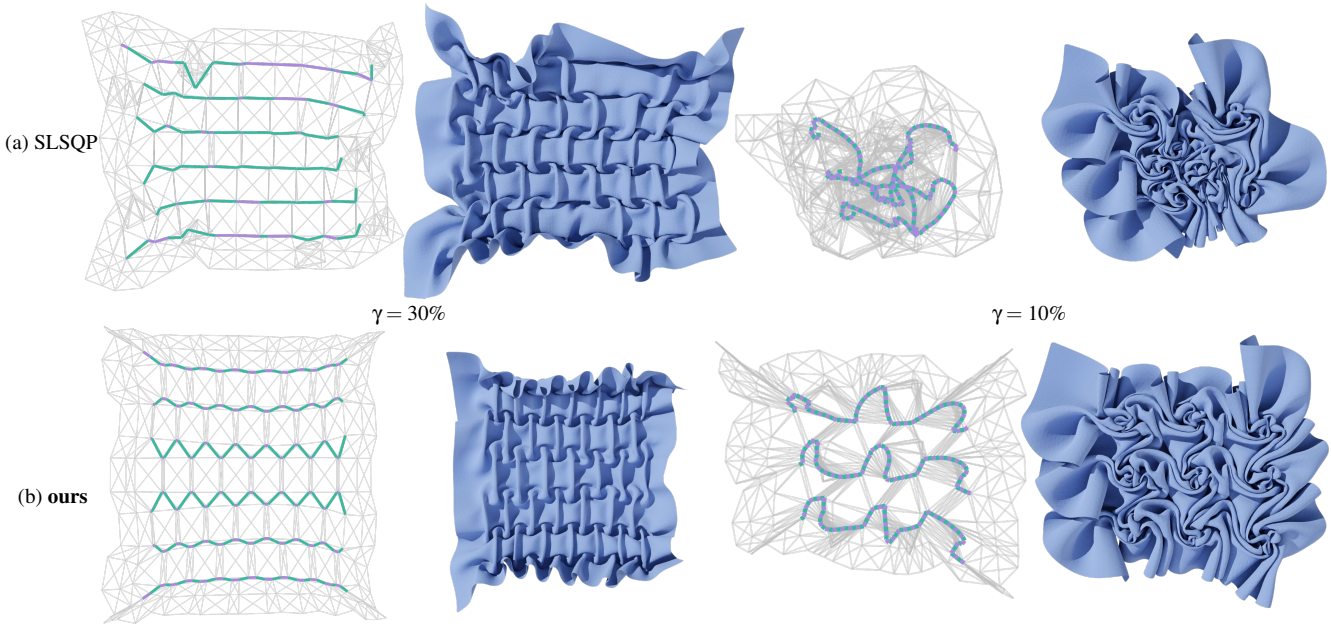

\begin{table}[!t]
    \caption{\new{We report the input complexity for the examples in \figref{fig:appx:convergence}, including \#stitching vertices $\vert\Vs\vert$, \#grid vertices $\vert \V \vert$, \#constraints  $N_c$ in Alg.~\ref{algo:2d-sim}, and shrinkage $\gamma$, and the runtime for 2D simulation.}}\label{tab:appx:sqp}
    \centering
\footnotesize
{\def\arraystretch{1.1}\tabcolsep=0.8em
\begin{tabular}{ccccccc}
\toprule[1pt]
\multirow{2}{*}{\begin{tabular}[c]{@{}c@{}}\itshape smocking\\ 
\itshape pattern\end{tabular}}  & \multicolumn{4}{c}{ \itshape complexity} & \multicolumn{2}{c}{ \itshape 2D sim.\ runtime (sec)} \\ \cmidrule[0.8pt](l){2-5}\cmidrule[0.8pt](l){6-7}
&   $\vert\Vs\vert$       & $\vert \V \vert$       & $N_c$    & $\gamma$    & SLSQP   & \textbf{ours}  \\ \midrule[1pt]
Fig.~\ref{fig:res:multi-row-patterns} (a) & 174 & 304 & 2562 & 30$\%$ & 161.5  & 2.2  \\
Fig.~\ref{fig:teaser} & 237 & 437 & 3716 & 10$\%$ & 3140.1  & 68.8    \\
\bottomrule[1pt]
\end{tabular}
}

\end{table}

\section{\new{SQP for 2D simulation}}\label{app:sqp}
An alternative solution to the problem in \eqnref{eq:mtd:2d:prob:all} is using off-the-shelf solvers, such as sequential quadratic programming (SQP)~\cite{Boggs1995sqp}, designed for constrained nonlinear optimization.
Here we compare Scipy's SLSQP solver~\cite{kraft1988slsqp,scipy2020} to our Algorithm~\ref{algo:2d-sim} in solving \eqnref{eq:mtd:2d:prob:all} on two different patterns: the \textsc{Curve} pattern shown in \figref{fig:teaser} and the pattern in \figref{fig:res:multi-row-patterns}~(a).

We report the problem complexity and runtime comparison in Table~\ref{tab:appx:sqp}, and show the corresponding qualitative comparison in \figref{fig:appx:sqp}.
It is worth noting that solving \eqnref{eq:mtd:2d:prob:all} is quite challenging, primarily due to the significantly greater number of constraints ($N_c$) compared to the number of variables ($\vert\V\vert$).
The SLSQP solver manages to find feasible solutions even when the initialization (planar fabric) does not fully satisfy all constraints. 
However, the solver still struggles with generating regular pleats given the complicated search space from the large number of constraints. This also justifies our choice of designing a specialized solver, leading to more accurate and visually pleasing results at a significantly reduced computational cost.

% \JR{add discussion about scipy solver, add table to report number of variables/constraints, runtime for slsqp and ours}
% \JR{add a figure showing the simulated results using sqp and our results for 2D simulation}

% \begin{figure}[t]
%     \centering
%     \begin{overpic}[trim=0cm 1.5cm 0cm 1cm,clip,width=1\linewidth,grid=false]{figures/appxC_rod_fabric_with_ours.pdf}
%     \put(19,35.2){\footnotesize (a) pattern}
%     \put(20,15.2){\footnotesize (b) input}
%     \put(67,43){\footnotesize (c) unfixed fabric}
%     \put(66,28){\footnotesize (d) left-fixed fabric}
%     \put(73,12){\footnotesize (e) \textbf{ours}}
%     \end{overpic}
%     \caption{We compared the results of smocking with rod-fabric interaction with ours.}
%     \label{fig:appx:rod_fabric}
% \end{figure}

\section{\new{Rod-fabric interaction modeling for smocking}}\label{app:rod-fab}
\begin{figure}[t]
    \centering
    \begin{overpic}[trim=0cm 0.95cm 0cm 1cm,clip,width=1\linewidth,grid=false]{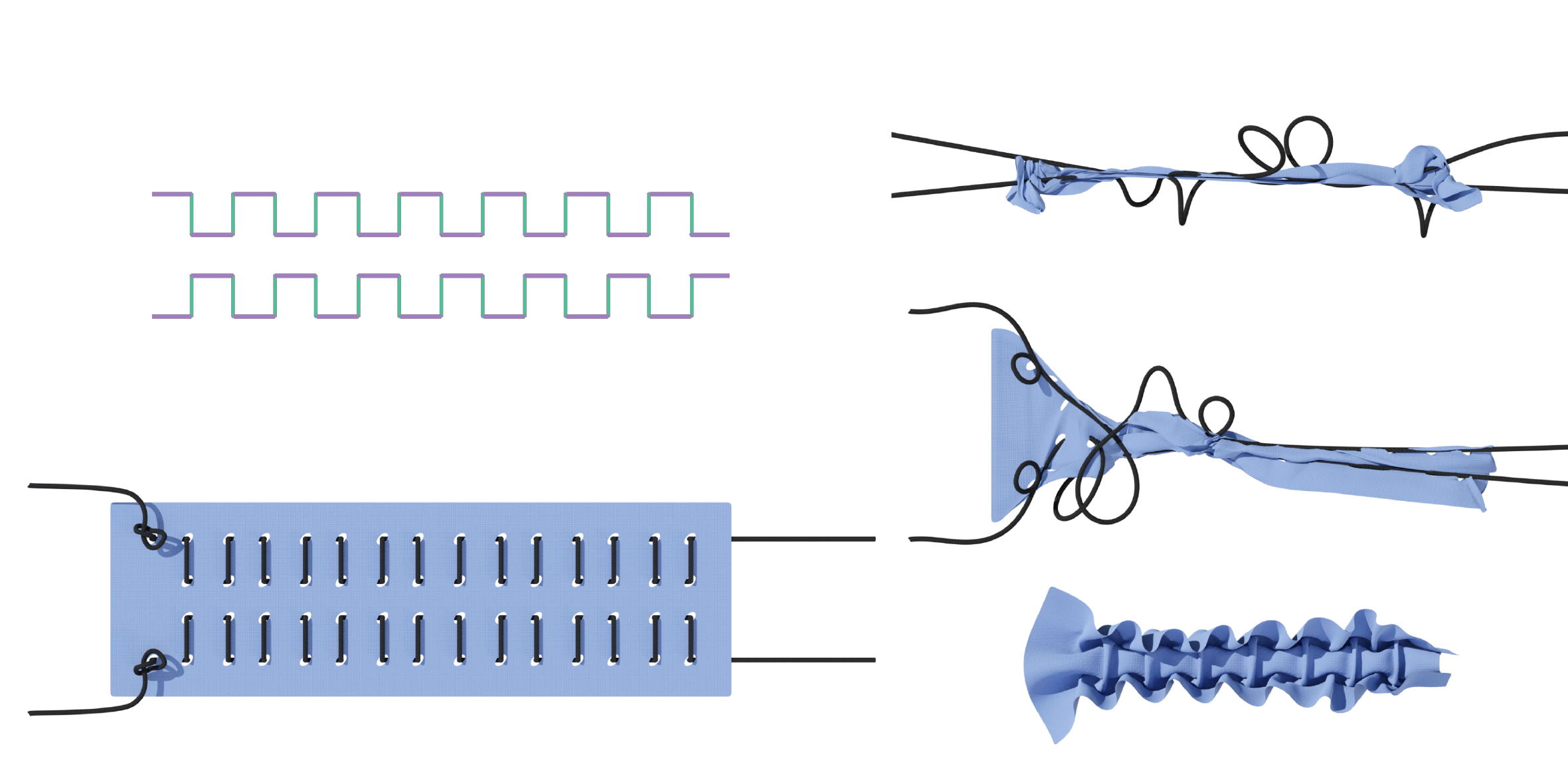}
    \put(19,37){\footnotesize (a) pattern}
    \put(15,17){\footnotesize (b) input (front side)}
    \put(67,41){\footnotesize (c) unfixed fabric}
    \put(66,26){\footnotesize (d) left-fixed fabric}
    \put(73,9.5){\footnotesize (e) \textbf{ours}}
    \end{overpic}
    \caption{\new{We compare the results of formulating smocking via rod-fabric interaction to our thread-fabric interaction.}}
    \label{fig:appx:rod_fabric}
\end{figure}
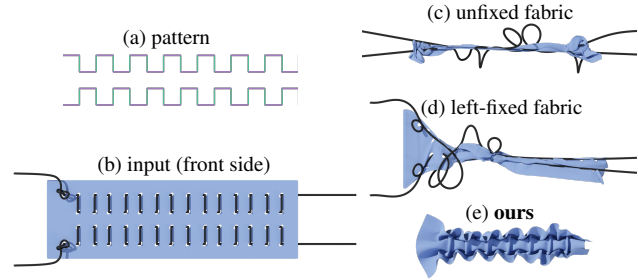

An alternative approach to simulating smocking involves treating the long threads as thin rods and exploring the interaction between the rods and fabric.
Specifically, we thread thin rods through the pre-drilled tiny holes in the fabric. 
The initial positioning of the rods aligns with the input smocking pattern, with additional height adjustments relative to the front and back stitches.
One end of the rods is secured with knots, while the other end is left free to simulate the pulling process in real fabrication, as shown in \figref{fig:appx:rod_fabric}~(b).

We encountered several non-trivial challenges when modeling the fabrication process: 
(1) The real fabrication process includes both pulling the threads and pushing the fabric in the opposite direction. It is unclear how to translate the pushing action of the fabric into the simulation process.
(2) Since the fabric undergoes shrinkage in both vertical and horizontal directions during the process,
setting up accurate fixed-boundary conditions for simulation involves unknown positions.
(3) Since multiple local minima exist, it is necessary to include a regularizer to find more regularly shaped pleats, which is challenging to formulate.

In all configurations we have explored, the final results are cluttered with irregular pleats, worse than the baselines we considered (i.e., Blender and C-IPC without any priors).
See \figref{fig:appx:rod_fabric}~(c) and \figref{fig:appx:rod_fabric}~(d) for two representative failure cases encountered.
Both experiments involve directing the free ends of the rods to move in a constant rightward direction. We model the pushing action of the fabric as a body force to the left applied across the entire fabric (setting (c)). In setting (d) we additionally fix the left side of the fabric.
The simulated results in these naive configurations prove unsatisfactory. 
In setting (c), despite applying a sufficiently strong body force, the fabric slides along the rods before the pleats form when the rods are pulled -- an occurrence that is unlikely in reality. 
To counter such sliding,  the left side of the fabric is fixed in setting (d), resulting in only marginally improved results.
Besides, both settings struggle to achieve a regular distribution of fabric shrinkage, showcasing the complexity of modeling the pulling action. 
Additionally, the overly cluttered regions induce complex (and mostly unnecessary) collision handling that requires significantly higher computational resources. 
These challenges emphasize the benefits of integrating geometric priors into the simulation, which leads to aesthetic pleats at significantly lower computational costs.

\end{document}